# Bright and Color-Tunable Single Layer Perovskite Host-Ionic Guest Light Emitting Electrochemical Cells

*Aditya Mishra[1], Stephen DiLuzio[2], Masoud Alahbakhshi[3], Austen C. Adams[4], Melanie H. Bowler[4], Qing Gu[3], Anvar A. Zakhidov[4,5,6*], Stefan Bernhard[2*], and Jason D. Slinker[1,4]**

[1]Department of Materials Science and Engineering, The University of Texas at Dallas, 800 West Campbell Road, Richardson, Texas 75080-3021, United States.

[2]Department of Chemistry, Carnegie Mellon University, 4400 Fifth Avenue, Pittsburgh, PA 15213

[3]Department of Electrical and Computer Engineering, The University of Texas at Dallas, 800 West Campbell Road, Richardson, Texas 75080-3021, United States.

[4]Department of Physics, The University of Texas at Dallas, 800 West Campbell Road, Richardson, Texas 75080-3021, United States.

[5]NanoTech Institute, The University of Texas at Dallas, 800 West Campbell Road, Richardson, Texas 75080-3021, United States.

[6]Laboratory of Advanced Solar Energy, NUST MISiS, Moscow, 119049, Russia

AUTHOR INFORMATION

**Corresponding Author**

*slinker@utdallas.edu, *bern@cmu.edu, *zakhidov@utdallas.edu

ABSTRACT: Perovskite light emitting devices have drawn considerable attention for their favorable optoelectronic properties. High carrier mobilities make perovskites excellent candidates as host materials in electroluminescent devices. To achieve high performance in a simple single layer device, we employed a $CsPbBr_3$ perovskite host and a novel ionic iridium complex guest along with a polyelectrolyte to demonstrate efficient light emitting electrochemical cells (PeLECs). For an optimal guest/host blend, 10600 cd/m$^2$ luminance at 11.6 cd/A and 9.04 Lm/W are achieved at 4.1 V. These devices showed voltage-dependent electroluminescence color proceeding from orange-red to green, facilitated by the reconfigurable ionic materials blend. Optimized devices exhibited stable operation under constant current driving, maintaining >630 cd/m$^2$ emission for 40 h. Our rationally-designed ionic guest at an optimal concentration of the host produced efficient (>90%) Förster energy transfer and improved thin film morphologies for high performance PeLEC operation enabled by ionic migration to interfaces.

Electroluminescent (EL) device technology is advancing to address commercial interests in flat panel displays, solid-state lighting, and diverse interfaces with smart technologies. Power efficiency, color tunability and low cost are key factors in selecting materials for these assorted applications. Inorganic light-emitting diodes (LEDs)[1-3] and organic light-emitting diodes (OLEDs)[4-6] have emerged for a number of applications, the former for efficient and robust operation and the latter for color purity and facile fabrication. At the intersection of these efforts, perovskite-based LEDs have exhibited peak brightnesses and efficiencies comparable to inorganic LEDs and the characteristic solution processing of OLEDs.[7-13] This impressive performance has led to significant interest in developing these materials for optoelectronic applications.

Facilitating this rise to prominence in EL devices, perovskites are also excellent solution-processable conductive materials, showing enhanced electron and hole lifetimes[14] with mobilities surpassing 10 $cm^2/V \cdot s$.[15-17] Utilizing this feature, a number of reports now describe the use of perovskites as hosts within LEDs (See Supplemental Information Table S1). Yao *et al.* achieved white emission at 350 $cd/m^2$ maximum luminance by using a $CsPbBr_xCl_{3-x}$ host with a poly[2-methoxy-5-(2-ethylhexyloxy)-1,4-phenylenevinylene] (MEH-PPV) guest together with hole and electron transport layers.[18] Matsushima *et al.* leveraged methylammonium lead chloride ($MAPbCl_3$) as a host layer in conjunction with a coumarin emitter in a multilayer architecture and achieved 8.3 cd/A current efficiency and 5000 $cd/m^2$ maximum luminance.[19] Chang *et al.* demonstrated color tuning with $MAPbBr_3$ and $MAPb(Br_{0.6}Cl_{0.4})_3$ hosts and orange-near infrared organic emitters with a 0.01 maximum external quantum efficiency.[20] Zhang *et al*. leveraged a $MAPbBr_3$ host with an $Ir(piq)_2acac$ emitter, where acac is acetylacetonato, and a 1,3,5-tris(1-phenyl-1*H*-benzimidazol-2-yl) benzene (TPBi) electron transport layer. Devices exhibited a maximum luminance of 742 $cd/m^2$ and a maximum current efficiency of 0.17 cd/A.[21] Cortecchia *et al*. reported the use of a two dimensional $NMA_2PbBr_4$ host, where NMA is 1-napthyl-methylammonium, and a manganese guest, achieving orange emission at 0.004% EQE.[22] These strategies have demonstrated that a perovskite can serve as a host. However, high performance has been elusive and has yet to be achieved in simple single layer devices. Maintaining phase compatability between the perovskite host and the guest has also proved to be difficult, contributing to the challenge of high performance.

Recent reports have demonstrated that perovskites can be leveraged as single-layer light-emitting electrochemical cells (PeLECs),[23-29] with emerging studies demonstrating high brightness[25,26,29] and long operational duration.[29] Light emitting electrochemical cells (LECs) utilize ionic motion to generate electrical double layers (EDLs) for high carrier injection and potential doping effects to achieve efficient operation in single layer devices. Recently, we observed that blending $CsPbBr_3$ with poly(ethylene oxide) (PEO), an electrolyte, and $LiPF_6$, an ionic additive, produces devices capable of supporting high luminance of ~15000 cd/m$^2$ and long lifetimes extrapolating to 6700 h at 100 cd/m$^2$.[25,29] Balancing these three components produced highly uniform and smooth films, enhanced thin film photoluminescence quantum yield, and optimal electrical double layer formation in PeLECs. Thus, PeLECs achieve enhanced functionality in a simple, single layer structure and accordingly present a new method for pursuing optimized host-guest effects, particularly if the guest emitter can be designed as ionic in nature.

Color tunability by voltage is an attractive function of LEDs and LECs, particularly guest/host systems. Voltage-tunable color can be achieved by multilayer structures,[30-34] blends involving metal complexes,[35-37] and OLED tandems[38-40] and a comparative table of their performance is shown in Table S2. Among these, LECs enable this color tuning functionality in a simple, single layer architecture. Furthermore, the unique properties of perovskite hosts—high mobility, ionically-reconfigurable and high emissivity—offer a promising landscape for bright and wide band color tuning in these minimalistic LEC architectures.

Here, to achieve high performance and color tunability in a PeLEC host-guest device, we designed and synthesized an ionic iridium complex guest for compatibility and functionality with a perovskite host. Bright, efficient, and stable operation is achieved in a simple single layer host-guest blend device. Furthermore, broad color tunability is observed, facilitated by the unique blend of ionically-reconfigurable materials. The basis of these effects are understood from computational, photophysical, morphological, electrochemical, and optoelectronic characterization and justified according to the interplay of the underlying physics and chemistry of these mixed ion-electron conductors.

## Results

**Requirements and rationale for guest emitter.** Cyclometallated Ir(III) complexes have been and continue to be of great interest as luminophoric materials for organic photoredox transformations,[41,42] solar energy conversion,[43,44] and organic light emitting devices.[36,45,46] Whereas Zhang *et al.* previously utilized a conventional *charge neutral* Ir complex as a guest in a perovskite host,[21] we employed an *ionic* complex to promote phase compatibility with the perovskite host and additional ionic migration functionality. Furthermore, the guest complex should strongly absorb where the perovskite emits and possess a smaller bandgap to promote efficient energy transfer. Finally, it should adopt a size that is comparable to the unit cell of $CsPbBr_3$, that is, with a diameter on the order of 1 nm. Accordingly, we designed and synthesized a novel ionic emitter, [Ir(*t*-Bupmtr)$_2$(bathophen)][$PF_6$], where *t*-Bupmtr is 4-(4-*tert*-butyl-phenyl)-1-methyl-1H-1,2,3-triazole and is bathophen 4,7-Diphenyl-1,10-phenanthroline, to meet these optical and morphological requirements. Further details of this molecule are presented in the Supplemental Information and Methods sections.

**Host-guest thin film optical properties.** Thin films of the $CsPbBr_3$ host, the [Ir(*t*-Bupmtr)$_2$(bathophen)][$PF_6$] guest, and host-guest blends of various concentrations of guest were prepared by spin coating to test the optical properties of each. In Fig. 1a, we present the photoluminescence (PL) of thin films of the host and guest as well as the absorption of the guest thin film. The $CsPbBr_3$ PL is centered at 523 nm, consistent with our prior measurements.[25,29] The PL peak of [Ir(*t*-Bupmtr)$_2$(bathophen)][$PF_6$] film is centered at 591 nm, consistent with our solution measurement (see Supporting information). The absorption of the guest Ir complex has significant overlap with the emission of the perovskite host, satisfying this targeted requirement for effective energy transfer. In Fig. 1b, the normalized film PL of the host and guest are presented, expressed as a function of wt% of the Ir complex guest, along with films of pure host and pure guest. All blends show significant emission from both $CsPbBr_3$ and [Ir(*t*-Bupmtr)$_2$(bathophen)][$PF_6$], but the ratio of emission between the Ir complex guest and the perovskite host varies with concentration. The ratio increases with guest weight fraction from 10% to 12%, remains high for 14% and subsequently decreases for higher concentrations of the guest. Presumably, the initial increase follows from establishing a sufficient concentration of guest to harvest the host excitions, and the subsequent loss of luminance follows from self-quenching effects of the guest.[47] Thus, steady state PL measurements suggest potential energy transfer between the perovskite host and Ir complex guest, and an optimal concentration of guest maximizes guest emission.

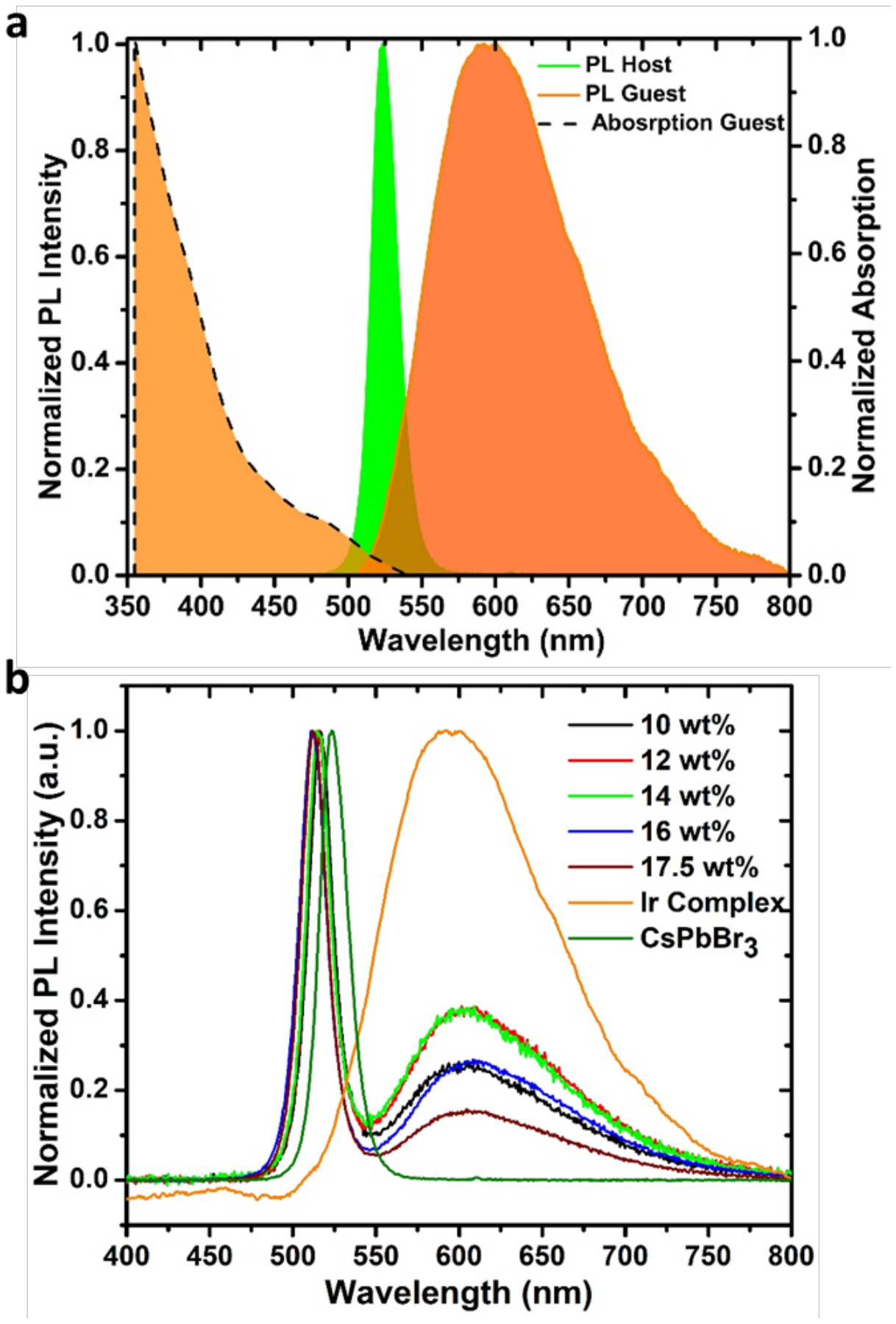

**Fig. 1 Thin film photoluminescence (PL) and absorbance spectra. a** Normalized PL of thin films of the $CsPbBr_3$ host and the [Ir(*t*-Bupmtr)$_2$(bathophen)][$PF_6$] guest, as well as thin film absorbance of the [Ir(*t*-Bupmtr)$_2$(bathophen)][$PF_6$] guest. **b** Normalized thin film PL of the $CsPbBr_3$ host, the [Ir(*t*-Bupmtr)$_2$(bathophen)][$PF_6$] guest, and host guest blends of various percent by weight guest.

**PeLEC fabrication, operation, and electroluminescence (EL).** PeLECs were constructed with a single layer semiconducting and electrolytic film using an ITO anode (with a poly(3,4-ethylenedioxythiophene) polystyrene sulfonate (PEDOT:PSS) modification layer to improve smoothness), a $CsPbBr_3$:[Ir(*t*-Bupmtr)$_2$(bathophen)][$PF_6$]:PEO perovskite composite (1:0.8 weight ratio CsPbBr3:PEO, various Ir complex weight fractions), and a LiF/Al cathode. With these blends, the devices functioned as PeLECs, as illustrated in Fig. 2a. Ions of the perovskite and Ir complex salt that are initially distributed uniformly through the film drift in response to an applied bias as facilitated by the PEO electrolyte. Cations drift toward the cathode, and anions toward the anode, leading to electric double layer (EDL) formation at each electrode. The EDLs contribute high electric fields at each contact that decrease the width of the potential barriers (doping) and enhance tunneling injection of electrons and holes, leading to exponentially increasing current. These injected carriers are then transported through the bulk and radiatively recombine as excitons. The band level diagram for the perovskite host, Ir complex guest, and electrode workfunctions are shown in Fig. 2b. Under this configuration, the [Ir(*t*-Bupmtr)$_2$(bathophen)][$PF_6$] is well aligned with the perovskite to exchange carriers with the $CsPbBr_3$ host. Interestingly, the energy levels derived from solution electrochemistry (See Supporting Information Figure S9) for the lowest unoccupied molecular orbital (LUMO) and highest occupied molecular orbital (HOMO) of the Ir complex guest lie just beyond those for the perovskite, making the guest electronic bandgap larger than the host and opposing the usual guest design rules for efficient energy transfer to guest. However, the strong spin-orbit effects of Ir facilitate faster rates of intersystem crossing (ISC),[47] allowing access to the triplet state of [Ir(*t*-Bupmtr)$_2$(bathophen)][$PF_6$] with a lower optical bandgap as illustrated in Fig. 2b by the intermediate level at −3.61 eV (obtained from the solid state PL). This smaller triplet optical bandgap supports triplet-triplet energy transfer from the perovskite host, facilitated by the heavy atoms (Pb) of the perovskite. This ISC relaxation process can likewise contribute to electron trapping and subsequent electrostatic hole capture, additionally contributing to the details of guest emission beyond the emission fraction via energy transfer from the host.[47]

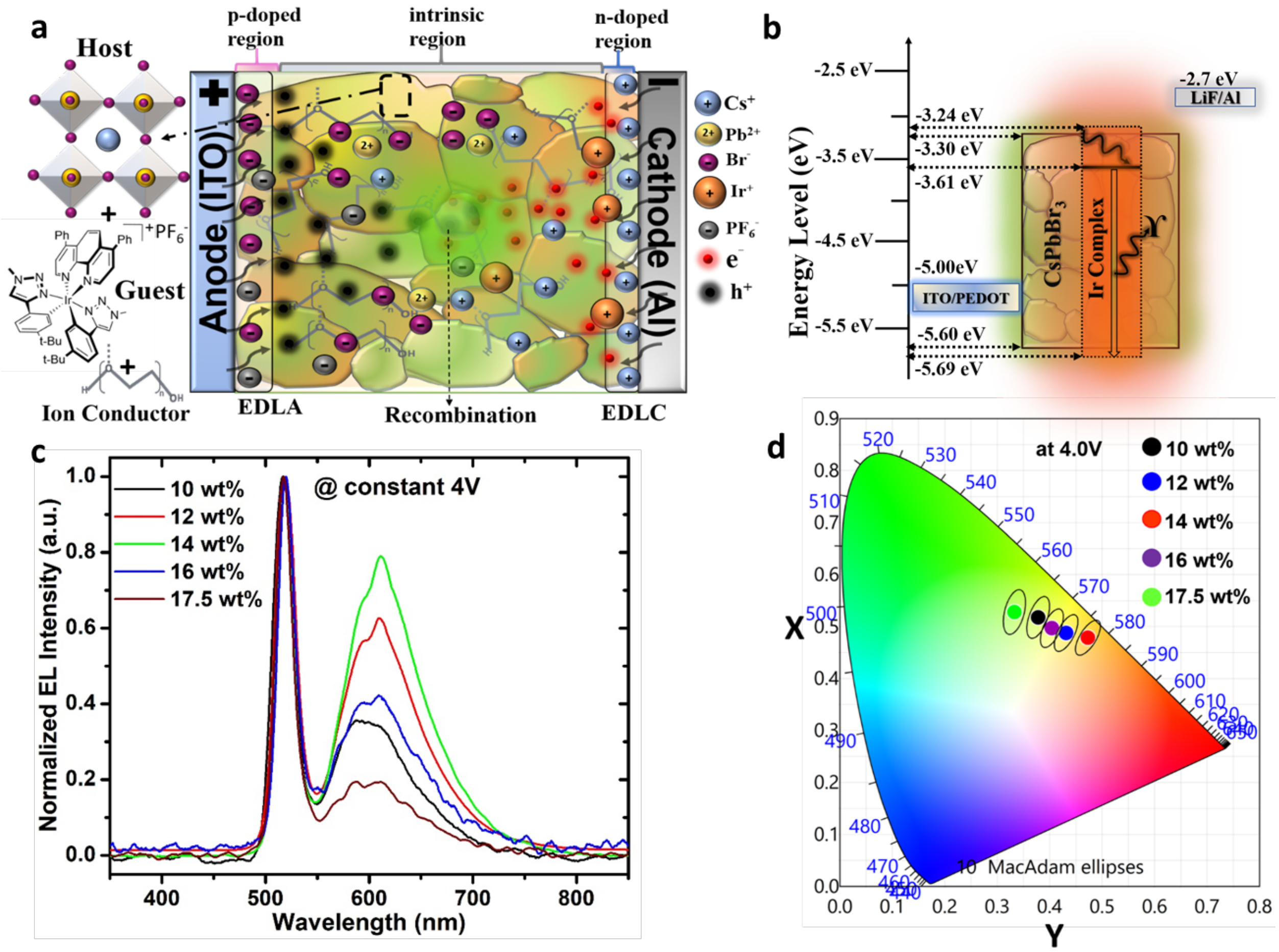

**Fig. 2 PeLEC operational mechanism, band level diagram, and EL characteristics. a** Illustration of the PeLEC device and operational mechanism. Thin films of blends of $CsPbBr_3$, [Ir(*t*-Bupmtr)$_2$(bathophen)][$PF_6$], and PEO are prepared between two electrodes. Upon application of a bias, ions redistribute to facilitate electron and hole injection and subseqeuent light emission. **b** Energy band level diagram for the $CsPbBr_3$ host, [Ir(*t*-Bupmtr)$_2$(bathophen)][$PF_6$] guest, ITO anode and LiF/Al cathode. Green and orange-red coloring represents the EL color from each component. **c** Electroluminescence (EL) spectra at 4.0 V for host-guest PeLECs with various weight fractions of guest. **d** International Commission on Illumination (CIE) coordinates for EL at 4.0 V for PeLECs with various weight fractions of guest.

The normalized EL spectra of these PeLECs of various weight ratios of $CsPbBr_3$ and [Ir(*t*-Bupmtr)$_2$(bathophen)][$PF_6$] under 4.0 V bias are presented in Fig. 2c. For comparision, the EL spectra of a $CsPbBr_3$ + PEO device and a pure [Ir(*t*-Bupmtr)$_2$(bathophen)][$PF_6$] LEC are presented in Supporting Information Figure S1. Similar to the thin film PL spectra, the EL fraction from the Ir complex initially increases with increasing guest concentration, reaches a maximum for the 14% guest blend, and then decreases for higher guest concentrations. This produces a shift in color from orange-red to green, as can be seen in the International Commission on Illumination (CIE) coordinates of Fig. 2d. Thus, the degree of charge and exciton trapping by the [Ir(*t*-Bupmtr)$_2$(bathophen)][$PF_6$] guest is concentration dependent and maximizes at 14%, corresponding to an 8.5% molar fraction of guest relative to the host.

**PeLEC operation as a function of voltage.** We investigated the effects of voltage sweeping on the operation of host-guest PeLECs and present the results in Fig. 3. Current density and luminance are plotted against voltage in Fig. 3a for host-guest PeLECs with various weight fractions of the guest. (For comparision, the performances of a $CsPbBr_3$ + PEO device and a pure [Ir(*t*-Bupmtr)$_2$(bathophen)][$PF_6$] LEC with voltage sweeping are presented in Supporting Information Figure S2). The host-guest devices achieve turn-on voltages (luminance >1 cd/m$^2$) between 2.2-2.8 V, with the lowest turn-on volage achieved for the optimal 14 wt% guest device. This low turn-on voltage near (or even smaller than) the bandgap of the material is a hallmark of LEC devices with highly mobile ions.[48-50] Peak current density and luminance are likewise maximized by the 14 wt% guest blend device, which reaches a maximum luminance of 10600 cd/m$^2$ at 4.1 V. Maximum current efficiency and maximum power efficiency are presented in Fig. 3b. The maximum efficiencies for each followed the trend 10% < 17.5% < 12% < 16% < 14%, peaking at 11.6 cd/A and 9.04 Lm/W for the 14 wt% guest device. Not shown, the maximum external quantum efficiency of this device was 3.81%. Close inspection also reveals that peak efficiency values were achieved at lower voltages for this 14% guest blend, contributing to the high power efficiency. The maximum luminance, current efficiency and power efficiency are all higher than those values previously reported for perovskite host devices[18-22] and 10X-100X higher than the results previously obtained for Ir(piq)$_2$acac PeroLEDs.[21] Current voltage cycling is also found to be more reversible with lower hysteresis for this blend, as seen in Figure S3 of the Supplemental Information, demonstrating more efficient ionic redistribution of the ionic guest.

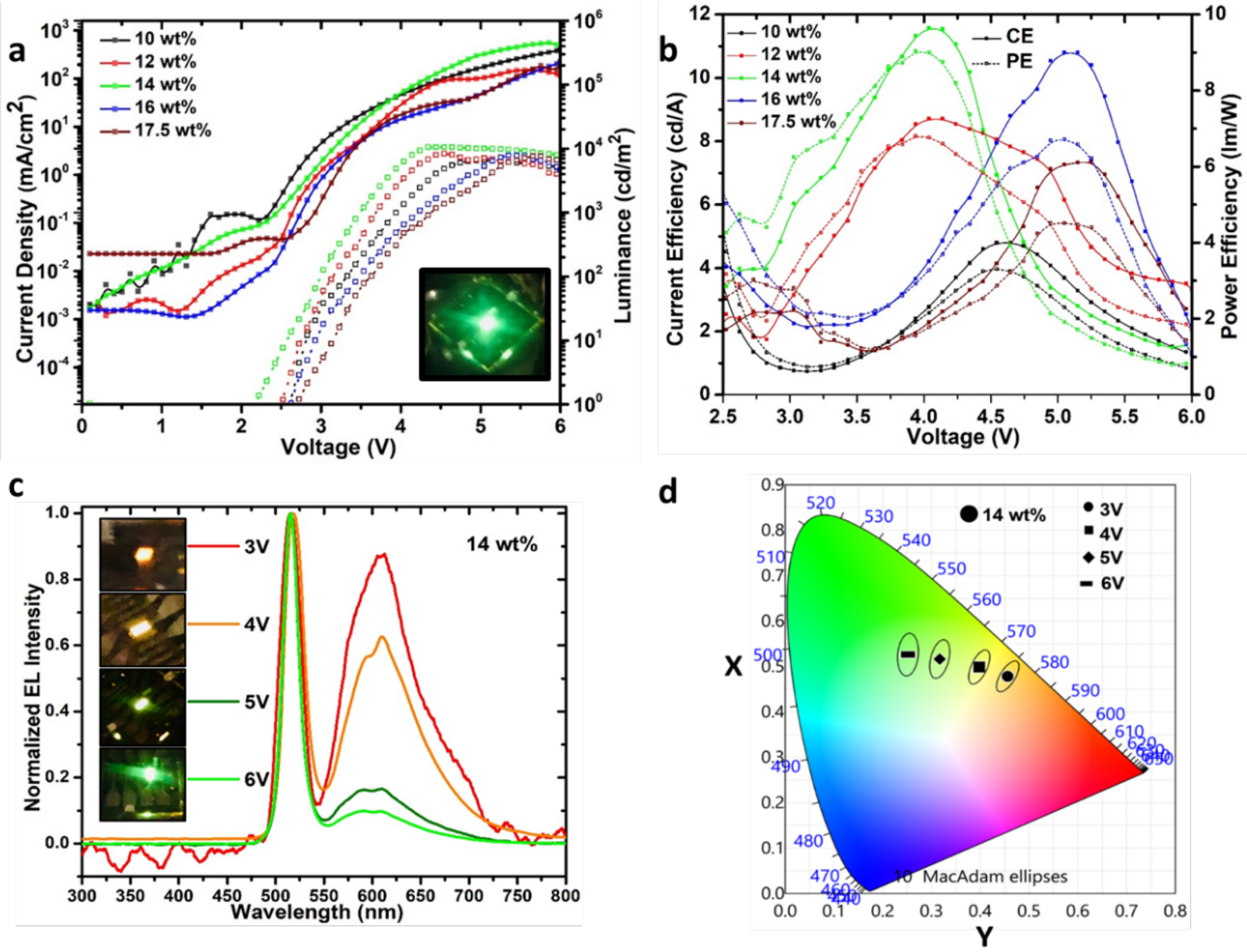

**Fig. 3 PeLEC operation as a function of voltage. a** Current density and Luminance vs. voltage for host-guest PeLECs with various weight fractions of guest. The inset is an example photograph of a device under 4 V operation. **b** Current efficiency and power efficency vs. voltage for host-guest PeLECs with various weight fractions of guest. **c** Electroluminesnce (EL) spectra for a host-guest PeLEC with 14 wt% guest at various voltages. **d** International Commission on Illumination (CIE) coordinates for EL of a host-guest PeLEC with 14 wt% guest at various voltages.

The EL exhibited broad color tunabilty with applied voltage. Fig. 3c shows the EL spectra of a 14 wt% [Ir(*t*-Bupmtr)$_2$(bathophen)][$PF_6$] PeLEC under various voltages along with images of the emissing devices, and Fig. 3d shows the associated CIE coordinates of these spectra. At 3 V, the emission appears orange and the EL spectra display nearly equal peak heights from the $CsPbBr_3$ host and [Ir(*t*-Bupmtr)$_2$(bathophen)][$PF_6$] guest. The significantly broader peak of the guest leads to a guest-dominant emission. At 4 V, emission appears yellow and the [Ir(*t*-

Bupmtr)$_2$(bathophen)][$PF_6$] contribution on the spectra is lowered relative to the 3 V case. At 5 V and 6 V the color is further shifted to yellow-green and green, respectively, with the [Ir(*t*-Bupmtr)$_2$(bathophen)][$PF_6$] peak successively dimished as voltage increases. The CIE coordinates are clearly shifted toward green and less saturated emission (away from the chart boundaries) as voltage increases. This effect can be understood from the saturation of the guest emissive state. At low voltages, most of the excitons are harvested by the Ir complex guest, and emission appears orange in conjunction with the [Ir(*t*-Bupmtr)$_2$(bathophen)][$PF_6$] PL spectra. Concomitantly, the CIE plot shows highly pure orange emission near the plot boundary. As voltage increases, these guest states become saturated with excitons due to the increased charge injection from ionic redistribution and subsequent charge trapping and energy transfer. The saturation of the guest leaves behind a remainder of excitons that radiatively decay in the $CsPbBr_3$ host leading to an increased fraction of green emission. Guest saturation may also lead to increased self-quenching effects. Similar voltage-dependent EL color tuning behavior is shown for other device blends (Figs. S4 and S5 of the Supplementary Information). Thus, a voltage-tunable color is observed, rangning from guest-dominated to host-dominated, following from saturation of guest states and potential quenching of guest excitations.

**PeLEC operation under constant current and constant voltage driving.** We recently demonstrated that PeLECs can achieve significantly long lifetimes when operated under constant current driving.[29] Constant current driving typically produces higher voltages in initial stages of LEC operation, inducing facile ionic redistribution during EDL formation.[51] The voltage then lowers as efficient charge injection decreases the device resistance (via potential p and n doping effects). The ability to operate at this lower voltage leads to greater stability of the device.

Fig. 4a shows the luminance versus time for host-guest PeLEC devices with various fractions of guest under constant current operation. In all cases, the luminance increases rapidly until the maxima are reached. Subsequently, luminance exponentially decreases for 4-5 h of operation, after which luminance remains in a steady state for ~18-20 h. Next, luminance again decreases exponentially for ~10-12 h, followed by a fourth region in which luminance increases. Clearly, device stability has a complex time behavior, and likely involves a combination of degradation and structural rearrangement. Among the PeLECs of various compositions, the 14 wt% blend produced the highest luminance throughout the operation as well as the greatest perceived stabilty. These optimized devices maintained >630 cd/m$^2$ emission for 40 h at 0.33 A/cm$^2$.

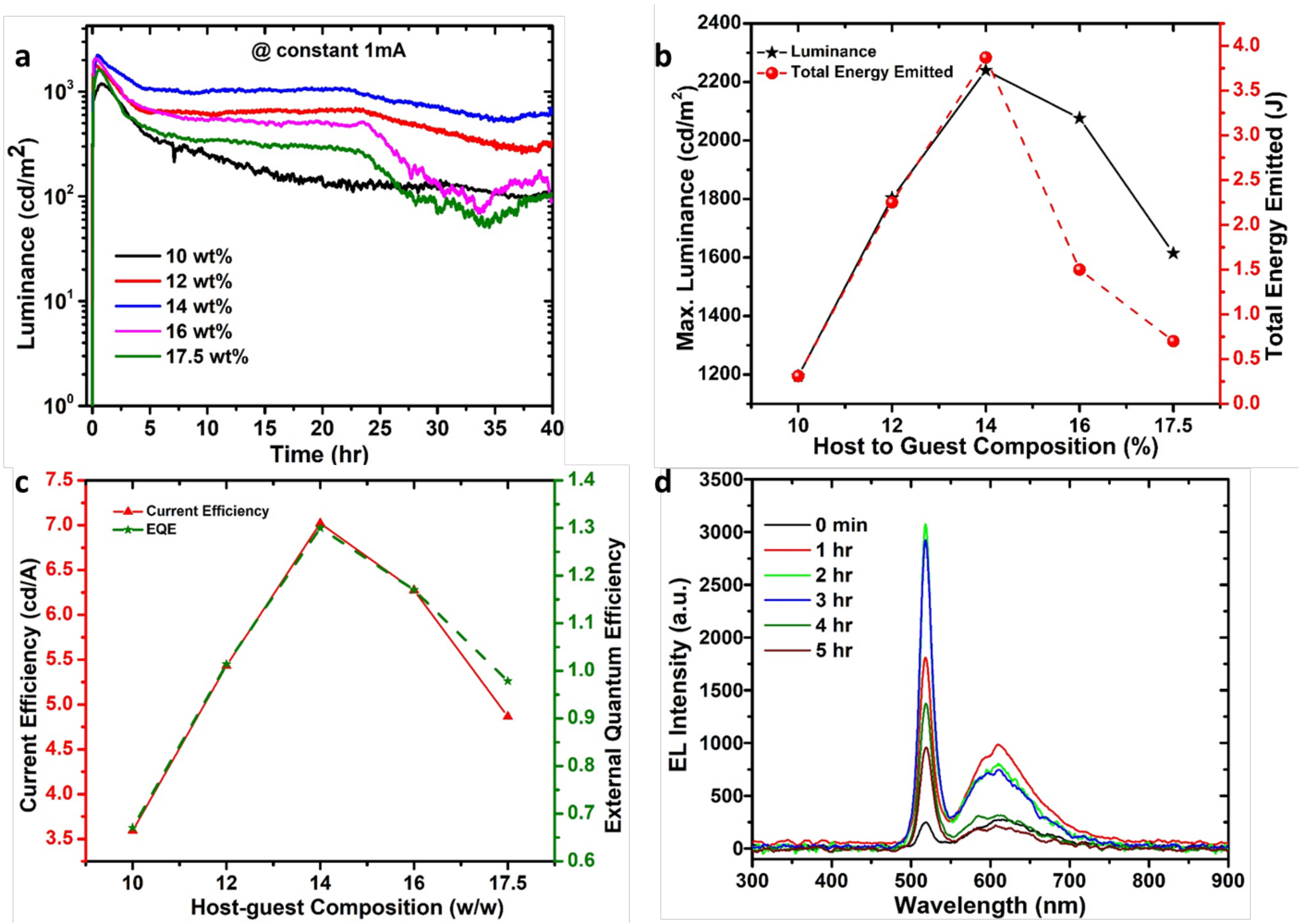

**Fig. 4 PeLEC operation under constant current and constant voltage driving. a** Luminance vs. time for host-guest PeLECs with various weight fractions of guest under 1.0 mA constant current driving (current density 0.33 mA/cm$^2$). **b** Maximum luminance and total emitted energy

versus guest concentration for host-guest PeLECs with various weight fractions of guest. **c** Current efficiency and external quantum efficiency vs. voltage for host-guest PeLECs with various weight fractions of guest. **d** Electroluminesnce (EL) spectra for a host-guest PeLEC with 14 wt% guest at various times under 4 V driving.

Fig. 4b summarizes the peak luminance and lifetime characteristics of these devices. The data reflect that the PeLEC with 14 wt% [Ir(*t*-Bupmtr)$_2$(bathophen)][PF$_6$] maximize both metrics. Peak luminance improves with concentration from 1200 cd/m$^2$ for the 10 wt% guest PeLEC to an optimized value of 2300 cd/m$^2$ for the 14 wt% PeLEC and subsequently decreases with contentration, ultimately to 1600 cd/m$^2$ for the 17.5 wt% device. To quantify lifetime, many metrics are inappropriate given the complex nature of the luminance versus time curves. Therefore, we analyzed PeLEC lifetime in a manner that avoids extrapolation. We calculated the total emitted energy ($E_{total}$) for each device by integrating the radiant flux curve up to the time where it decays to 1/5 of maximum.[52,53] This metric is also strongly peaked at 14 wt% guest at 3.8 J for a 3 mm$^2$ active area device (total emitted energy density $U_{total}$ = 1.3 J mm$^{-2}$), more than two times higher than all other blends.

To further understand the nature of device operation and degradation, in Figure S6 of the Supplementary Information we present the voltage versus time obtained from these PeLECs under 1.0 mA constant current driving. In all cases, the operational voltage begins at a maximum value between 3-4 V and slowly lowers over time, consistent with improved charge injection and potential doping effects from slow ionic redistribution and EDL formation. The 14 wt% guest PeLEC shows a substantially lower voltage that decreases to 2.3 V and remains steady through operation. The other device blends exhibit voltages that are higher and less stable with time. Thus, an optimized concentration of the guest leads to favorably low voltage operation in support of higher stabilty devices.

In Figure 4c the maximum current efficiency and external quantum efficiency from 1.0 mA constant current driving are presented as a function of guest concentration. See Supplementary Information Figure S7 for the power efficiency curve. As for the peak luminance and lifetime data, these maximum efficiencies are strongly peaked for the 14 wt% guest devices. For this optimal blend, the current efficiency peaked at 7.0 cd/A, the external quantum efficiency achieved 1.2%, and the power efficency attained 8.0 Lm/W.

To carefully explore the time dependence of ionic redistribution on chromaticity of the EL spectra, we tested the 14 wt% guest PeLEC under constant voltage as shown in Fig. 4d. Devices begin with strong orange emission and a proportionally larger contribution from the [Ir(*t*-Bupmtr)$_2$(bathophen)][$PF_6$] guest relative to the $CsPbBr_3$ host. As operational time proceeds to 1 h, luminance increases, with both guest and host peaks increasing in accordance with EDL formation (and thus better injection). However, at the 2 h and 3 h time points, the 523 nm host peak increases, while the 591 nm guest peak decreases. As with sweeping to higher voltages, this suggests that guest states become saturated and partially quenched as injected current increases in time, leading to enhanced exciton formation and radiative decay in the host. At later time points both peaks decrease. The dynamics of intensity ratios of host and guest fractions in time are plotted in Supplementary Information Figure S8, showing that EL initiates with 77% of the emission from the guest, drops linearly to 45% guest emission over two hours, and then remains somewhat stable. This unusual color change with time in these ionically reconfigurable materials is caused by the interplay of ionic redistribution, charge injection, doping, and quenching effects. Such effects can be controlled with variable voltage operation.

**Time resolved PL spectroscopy of host-guest films.** To determine the propensity for energy transfer in host-guest blends, we measured the time resolved PL of thin films of host, guest, and host-guest blends, as shown in Fig. 5. The PL intensity versus time ($I(t)$) of the perovskite host and the Ir complex guest both exhibited biexponential behavior of the form

$$I(t) = A_1 e^{\frac{-t}{\tau_1}} + A_2 e^{\frac{-t}{\tau_2}}$$

with time constants $\tau_1$ and $\tau_2$ on the nanosecond timescale. For $CsPbBr_3$, $\tau_1$ was 10.7 ns, and $\tau_2$ was 93.5 ns, and [Ir(*t*-Bupmtr)$_2$(bathophen)][$PF_6$] had similar behavior, with $\tau_1 = 6.2$ ns, and $\tau_2 = 58.1$ ns. By contrast, host-guest blended films (Fig. 3c) showed biexponential decay on the order of hundreds of picoseconds, considerably faster than the dynamics of either host or guest. Using the simple approximation for the efficiency of Förster energy transfer ($E_T$):

$$E_T = 1 - \frac{\tau_{HG}}{\tau_H}$$

where $\tau_{HG}$ is the lifetime of the host-guest system and $\tau_H$ is the lifetime of the pure host. With this approximation and using the fast time components of each, $E_T$ is estimated to be 92%-97% for the blended films. Thus, the rationally-designed [Ir(*t*-Bupmtr)$_2$(bathophen)][$PF_6$] guest shows excellent energy harvesting from the perovskite host.

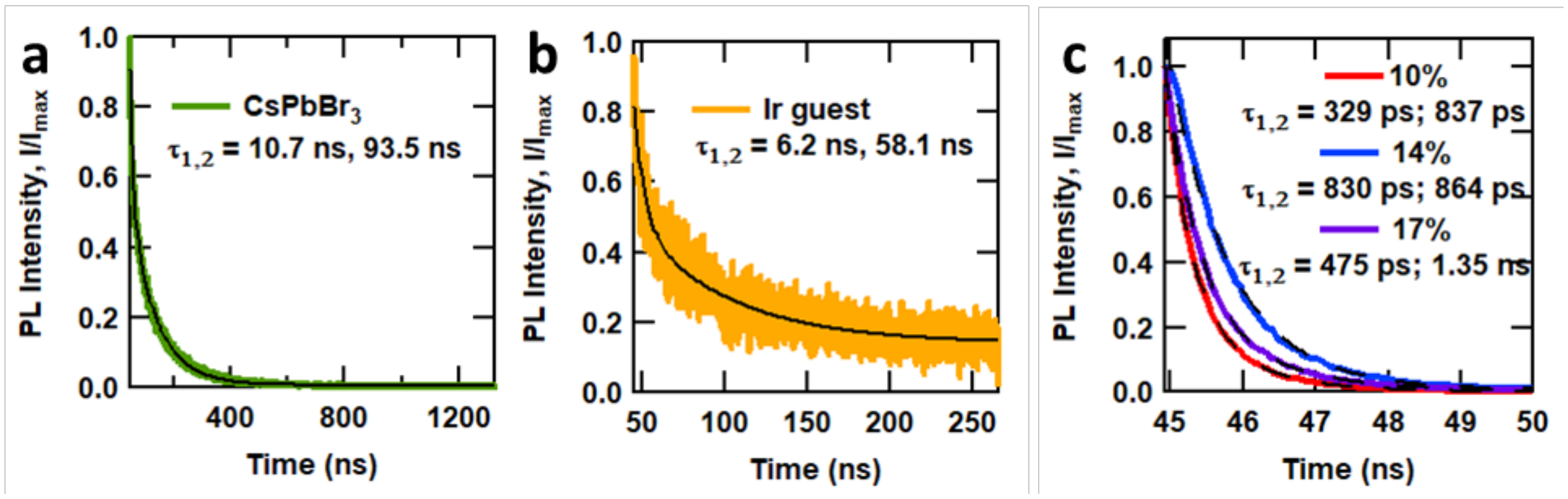

**Fig. 5 Time resolved PL spectroscopy of host, guest, and host-guest films. a** Time resolved PL of a $CsPbBr_3$ film. **b** Time resolved PL of a [Ir(*t*-Bupmtr)$_2$(bathophen)][$PF_6$]. **c** Time resolved PL from films of blends of $CsPbBr_3$ and [Ir(*t*-Bupmtr)$_2$(bathophen)][$PF_6$] expressed as various weight percentages of [Ir(*t*-Bupmtr)$_2$(bathophen)][$PF_6$].

**Morphology of host-guest films.** To gain further insight on the improvement of PeLEC performance for a distinct concentration of guest, the morphology of blends of $CsPbBr_3$, PEO, and various fractions of [Ir(*t*-Bupmtr)$_2$(bathophen)][PF$_6$] were studied with scanning electron microscopy (SEM), as shown in Fig. 6. SEM of the $CsPbBr_3$ + PEO film reveals that films are highly structured, with dense, polydisperse crystals ranging from 40 nm to 150 nm in diameter. The addition of the Ir complex guest leads to smoother films. The film with $CsPbBr_3$, PEO and 10 wt% [Ir(*t*-Bupmtr)$_2$(bathophen)][PF$_6$] shows a polycrystalline morphology with a lower density of crystals of reduced diameter ranging from 10-100 nm. The film with $CsPbBr_3$, PEO and 14 wt% [Ir(*t*-Bupmtr)$_2$(bathophen)][PF$_6$] appears smooth with a low density of crystals on the surface. The film with 17.5 wt% Ir complex was similar to the 14 wt% case but exhibited contrasted domains potentially indicative of phase separation. Thus, it appears that an optimized fraction of the guest improves thin film morphology. This stands in contrast to the prior study of a charge neutral Ir complex in a perovskite film, which showed significant phase separation with the charge neutral complex employed.[21] Also, whereas in the study of Zhang *et al.* the guest was cast on a pre-fabricated perovskite film with a chlorobenzene anti-sovent,[21] our host and guest (together with PEO) were mixed in solution and spin coated simultaneously from a common solvent. Here, the optimized fraction of cationic Ir complex improves thin film morphology with the ionic perovskite and PEO polyelectrolyte.

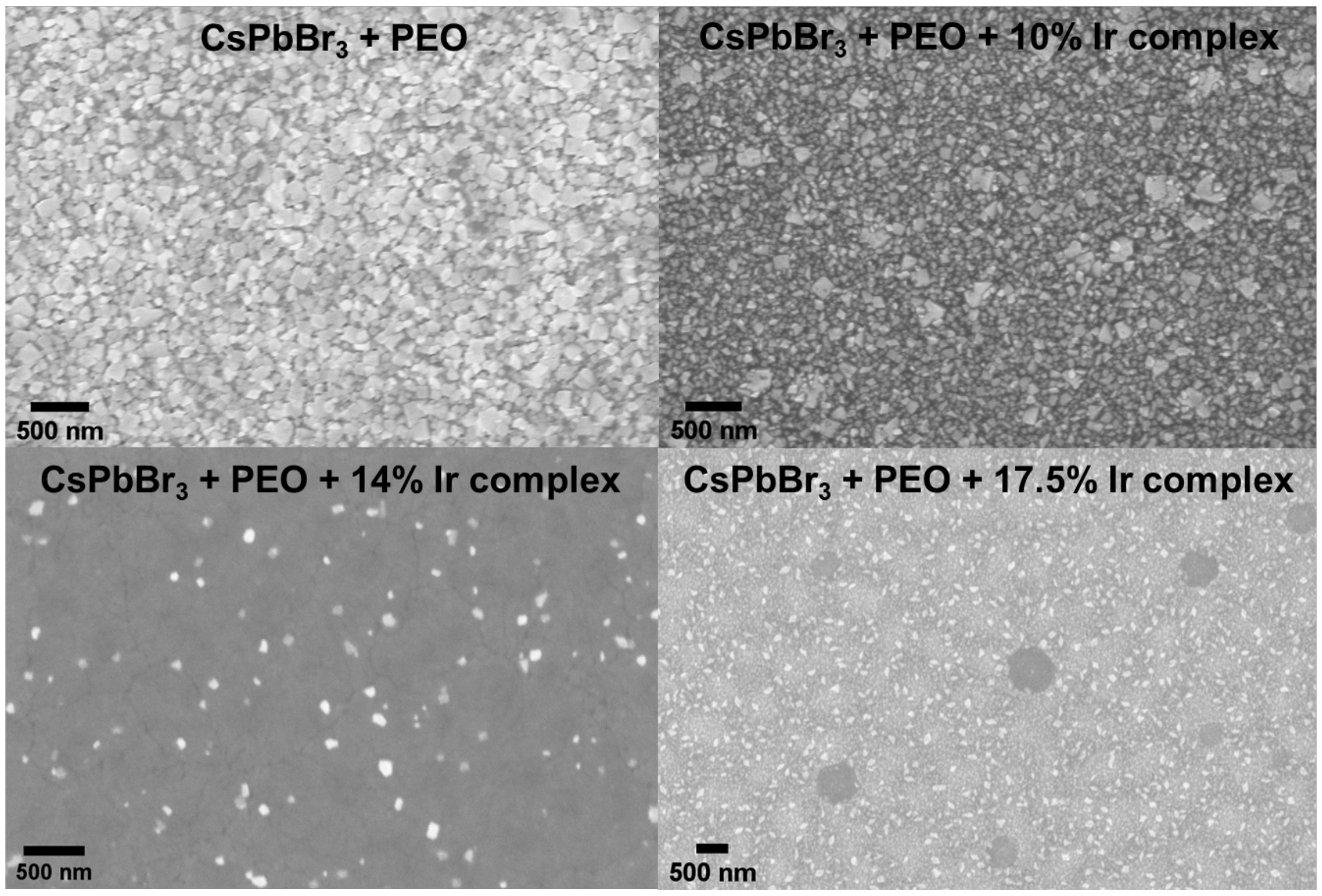

**Fig. 6 Morphology of host-guest films.** Scanning electron microscopy of thin films of $CsPbBr_3$ + PEO and $CsPbBr_3$ + PEO + various fractions of [Ir(*t*-Bupmtr)$_2$(bathophen)][$PF_6$].

## Discussion

Efficient, bright, and long-lasting PeLECs were successfully demonstrated from single layer devices using a perovskite host and an ionic Ir complex guest. The ionic complex [Ir(*t*-Bupmtr)$_2$(bathophen)][$PF_6$] was designed for optical, electrical, and morphological interface as a guest with a $CsPbBr_3$ host. Spectral overlap and efficient energy transfer was confirmed with steady-state and time resolved PL spectroscopy. Thin film PeLECs based on blends of [Ir(*t*-Bupmtr)$_2$(bathophen)][$PF_6$], $CsPbBr_3$, and PEO exhibited maximum luminance (10600 cd/m$^2$), current efficiency (11.6 cd/A) and power efficiency (9.04 Lm/W) from an optimal blend, all beyond metrics reported for perovskite host devices. Constant current driving revealed operation in excess of >630 cd/m$^2$ emission for 40 h at 0.33 A/cm$^2$. Moreover color tunability by voltage is demonstrated for simple single layer architecture of composite LEC due to ionic nature of guest. These results demonstrated the potential for utilizing the host-guest strategy in PeLECs with perovskite hosts and rationally-designed emitters that are ionic in nature.

## Methods

**Materials:** Lead (II) bromide ($PbBr_2$; 99.99% trace metal basis), Cesium bromide (CsBr; 99.99%) and Polyethylene Oxide (PEO; M.W. > 5,000,000) were all purchased from Alfa Aesar. Lithium Hexafluorophosphate ($LiPF_6$; 99.99%) and Dimethyl Sulfoxide (DMSO; anhydrous > 99.9 % ) were purchased from Sigma Aldrich. [Ir(*t*-Bupmtr)$_2$(bathophen)][$PF_6$] (guest material) was synthesized as described below.

**Synthesis of 4-(4-(*tert*-butyl)phenyl)-1-methyl-1H-1,2,3-triazole (4tBupmtr):** 2 mL of 4-*tert*-Butylphenylacetylene was reacted with 724.3 μL of Iodomethane (1: 1.05 molar ratio) through the "click" chemistry approach. Sodium Azide, Sodium Carbonate, CuI, and Sodium Ascorbate were added in 1: 1.1, 1: 1, 1: 0.025, and 1: 0.5 molar ratios, respectively. The reaction was left to mix overnight in a 3: 1 molar ratio of dimethylformamide/water. Afterwards, a 0.1 M aqueous solution of Tetrasodium EDTA was added and the resulting precipitate was filtered and dried. The filtrated was recrystallized with ethyl acetate and hexanes. $^{1}$H NMR (500 MHz, Chloroform-*d*) δ 7.80 – 7.73 (m, 2H), 7.72 (s, 1H), 7.49 – 7.43 (m, 2H), 4.15 (s, 3H), 1.36 (s, 9H). $^{13}$C NMR (126 MHz, Chloroform-*d*) δ 151.19, 147.98, 127.80, 125.72, 125.43, 120.34, 36.65, 34.63, 31.27. MS (m/z ESI, ACN) calc: 215.29 g/mol, found 216.149. 53.7% Yield.

**Synthesis of Tetrakis-(4-(4-(*tert*-butyl)phenyl)-1-methyl-1H-1,2,3-triazole)-μ-(dichloro)diiridium(III) Dimer ([IrCl(4tBu-pmtr)$_2$]$_2$).** 250 mg of 4tBu-pmtr was reacted with 217.2 mg of $IrCl_3$*$H_2O$, purchased from J&J materials, in 2-ethoxyethanol and water in a 3:1 v/v ratio at 125 ºC overnight in air. Addition of deionized water after cooling precipitated a yellow solid. The resulting precipitate was filtered and dried with diethyl ether. (55.1% Yield).

**Synthesis of [Ir(4tBu-pmtr)$_2$(bathophenanthroline)][PF$_6$].** 200 mg of [IrCl(4tBu-pmtr)$_2$]$_2$ was reacted with 114.83 mg of Bathophenanthroline (1: 2.1 molar ratio) in Ethylene Glycol overnight at 150 ºC in air. An orange precipitate crashed out after addition of a 0.1 M aqueous $K^+$ $PF_6^-$ solution. After 30 minutes of stirring, the solution was diluted with deionized water, filtered, and dried with diethyl ether. The compound was recrystallized though the vapor diffusion method with acetonitrile/diethyl ether. $^{1}$H NMR (500 MHz, Chloroform-*d*) δ 8.45 (d, *J* = 5.3 Hz, 2H), 8.11 (s, 2H), 7.84 (s, 2H), 7.68 (d, *J* = 5.3 Hz, 2H), 7.60 – 7.54 (m, 10H), 7.39 (d, *J* = 8.0 Hz, 2H), 7.01 (dd, *J* = 8.0, 2.0 Hz, 2H), 6.39 (d, *J* = 1.9 Hz, 2H), 4.00 (s, 6H), 1.16 (s, 18H). $^{13}$C NMR (126 MHz, Chloroform-*d*) δ 157.83, 151.23, 151.16, 149.98, 148.53, 145.53, 135.86, 132.82, 129.70, 129.65, 129.10, 128.95, 128.75, 126.10, 125.73, 121.97, 119.88, 119.20, 38.28, 34.54, 31.34. MS (m/z ESI, ACN) calc: 952.841 [M-$PF_6^-$], found 953.360 [M-$PF_6^-$]. (58.5% Yield).

**Host and Guest Solution Preparation.** The $CsPbB_3$ precursor solution was prepared by dissolving $PbBr_2$: CsBr (1:1.5 molar ratio) in DMSO and kept overnight for dissolution. PEO (10mg/ml) were prepared in DMSO solution. Host precursor solution was prepared by mixing $CsPbBr_3$ and PEO solution in 100:80 weight ratio. Ir-Complex (guest) precursor solution was prepared by dissolving Ir-Complex in DMSO by different weight ratios (10%, 12%, 14%, 16% and 17.5%) and kept overnight for complete dissolution. The final host-guest solution ($CsPbBr_3$ +PEO+ Ir-Complex) was prepared by mixing guest precursor solution with different weight ratio of Ir-Complex into host precursor solution.

**Device Fabrication.** The overall host-guest PeLEC device architecture is as follows: ITO/PEDOT:PSS/ active layer/ LiF/Al. The active layer consisted of $CsPbBr_3$, PEO and different concentrations of Ir-Complex. Prepatterned indium tin oxide (purchased from Thin Film Devices, Anaheim, CA) were cleaned in a sequence of non-ionic detergent wash, water bath sonication, and UV ozone treatment. Aqueous poly(3,4-ethylenedioxythiophene):polystyrene sulfonate (PEDOT:PSS) solutions (1.3−1.7%, Clevios AI 4083) were filtered through a 0.45 μm GHP filter and then spin-coated to obtain a ∼20 nm thick film on the ITO-coated glass substrates. These films were subsequently annealed at 100 °C for 10 minutes in a dry $N_2$ filled glovebox. The prepared active layer precursor solution was spin casted onto PEDOT: PSS at 1500rpm followed by vacuum treatment for 90 seconds and then thermally annealed at 75 °C for 5 minutes. The active layer thicknesses were generally 120-130 nm. To deposit the top electrode, samples were transferred to a vacuum chamber, and 20 Å LiF and 800 Å Al were deposited using a shadow mask that defined 12 devices per substrate, each with a 3 $mm^2$ device area.

**Device Testing.** L-I-V measurements were obtained with a 760D electrochemical analyzer from CH Instruments (Austin, TX) and a calibrated Labsphere integrating sphere with thermoelectric cooled Si detector. The EL spectra were measured by an Ocean Optics Model 0 Jaz Spectrometer. The electrical and radiant flux characteristics for lifetime study were obtained with a custom multiplexer testing station capable of measuring 16 light emitting devices simultaneously. In brief, this instrument served as a current or voltage source and measuring unit and captured radiant flux with a calibrated Hamamatsu photodiode (S2387-1010R) for each device.

**Steady State Photoluminescence Measurements.** The PL measurements were taken using CW 405nm laser diode to excite the thin films coated on glass substrate. The emission from samples were collected using an Ocean Optics QE65000 spectrometer coupled to a fiber optic cable. To block the laser excitation signal, a 450nm longpass dielectric filter was used in front of fiber optic cable. All measurements were done at the room temperature with 45% relative humidity.

**Time Resolved Photoluminescence Measurements.** Time resolved PL measurements were carried out using a microscope-based time-resolved PL system based on an Olympus IX71 inverted microscope. Samples were excited at 405 nm (300 fs) generated by frequency-doubling the output of a Ti:Sapph pulsed laser (MIRA 900F) focused on the sample surface using a NA = 0.9 objective. The emission, collected using the same objective, was passed through a spectrometer to a silicon avalanche photodiode (MicroPhotonDevices). Time-resolved PL was measured using time-correlated single photon counting (PicoQuant 300 GmbH).

**Scanning Electron Microscopy (SEM).** High resolution secondary electron SEM images were obtained using a Zeiss Supra-40 SEM using an in-lens detector at an accelerating voltage of 10kV. The taken images were analyzed by ImageJ software.

**Data availability**

The authors decalre that the main data supporting the findings of this study are contained within the paper. All other relevant data are available from the corresponding author upon reasonable request.

## Acknowledgements

J.D.S. acknowledges support from the National Science Foundation (ECCS 1906505). Q.G. acknowledges support from the Welch Foundation (AT-1992-20190330). A.Z. acknowledges support from the Welch Foundation (AT-1617) and from the Ministry of Education and Science of the Russian Federation (14.Y26.31.0010). S.B. acknowledges support by the US NSF (CHE-1764353).

## Author Contributions

A.M., S.D. and M. A. contributed equally to this work. S.D. prepared the Ir complex and characterized its photophysical and electrochemical properties. M.A. prepared solutions, performed PL study and SEM analysis. A.M., A.C.A., and M.H.B. prepared films, measured EL spectra, performed L-I-V analysis, and performed constant current study. J.D.S., S.B., A.M., S.D. and M.A. analyzed the data and prepared the draft, and Q.Z., and A.Z. helped revise the draft.

## Competing Interests

A.Z. is the president and CTO of Solarno, Inc., an early-stage technology development company interested in the commercialization of emerging nanomaterials, particularly carbon nanotubes, for heating, solar energy conversion, supercapacitors, and lighting applications.

## Additional Information

**Supplementary Information** is available free of charge at

**Materials and Correspondence**

Correspondence and requests for materials should be addressed to J.D.S. and S.B.

# Supplementary Information

# Bright and Color-Tunable Single Layer Perovskite Host-Ionic Guest Light Emitting Electrochemical Cells

*Aditya Mishra, Stephen DiLuzio, Masoud Alahbakhshi, Austen C. Adams, Melanie H. Bowler, Qing Gu, and Anvar A. Zakhidov, Stefan Bernhard, Jason D. Slinker**

Table S1. Comparative performances of the best-in-class perovskite host LECs. References coincide with numbered references of the main text.

| Device structure | Single or multilayer | Guest Emitter | Max Luminescence (cd/m2) | Max Current Efficiency (cd/A) | Max EQE (%) | Power Efficiency (Lm/W) | Ref |
|---|---|---|---|---|---|---|---|
| **ITO/ $NiO_x$ / $CsPbBr_xXI_{3-x}$:MEH-PPV/ TPBi/ LiF/ Al** | M (3 layers) | Conjugated polymer | 150 | - | - | - | 18 |
| **ITO/ HAT-CN/ Tris-PCz / $MAPbCl_3$:Coumarin/ T2T /BPy-TP2 / LiF/ AL** | M (5 layers) | Dye | 5000 | 8.32 | 2.02 | 8.44 | 19 |
| **ITO/ PEDOT:PSS/ $MAPbBr_3$:Organic Compound 1/ PBD/ LiF/ Al** | M (3 layers) | Red organic compound | 8 | 0.005 | 0.01 | - | 20 |
| **ITO/ PEDOT:PSS/ $MAPb(Br_{0.6}Cl_{0.4})_3$:Organic Compound 2/ PBD/ LiF/ Al** | | | 14 | 0.0025 | 0.005 | | |
| **ITO/ PEDOT:PSS/ $MAPbBr_3$:$Ir(piq)_2acac$/ TPBi/ LiF/ Al** | M (3 layers) | Phosphorscent | 742 | 0.17 | - | - | 21 |
| **ITO/ PEDOT:PSS/ PolyTPD/ $NMA_2PbBr_4$:Mn/ TPBi/ Ba/ Al** | M (4 layers) | Metallic doping | - | - | 0.004 | - | 22 |
| **Our Work** | Single layer | Iridium Complex | 10600 | 11.6 | 3.81 | 9.04 | |

Table S2. Color Tunable LED (LEC) devices, Mechanism of Voltage tuning of Chromaticity and performances of the perovskite host/guest LECs. References coincide with numbered references of the main text.

| Device structure | Multilayer or Single | Emitter Layer | Color Tuning Range and FWHM | Chromaticity Tuning | Mechanism of tuning | Voltage range (v) | Max Lum ($cd/m^2$) | Max EQE (%) | Ref |
|---|---|---|---|---|---|---|---|---|---|
| **ITO/PEDOT:PSS/ $CsPbBr_3$:AVAB/Red QD/blue QD/ZnO NPs/ Al** | M (5 layers) | Cd based Quantum dots | 460-520-624 nm FWHM: 24.4 – 20 – 31 nm | G-B: (0.1174, 0.7679) to (0.1454, 0.0464) R-G: (0.6791, 0.3198) to (0.1772, 0.7269) | The shift of main recombination zone | G-B: 3 - 7.5 G-R: 2.1 - 5.8 G-R-B: 8.5 - 14 | 11700 @ 7.4 v | - | 30 |
| **FTO/ / TiO2 / $MAPbI_{3-x}Cl_x$/ PbS:CdS QD/Spiro-OMeTAD / Au** | M (4 layers Bilayer emitter) | PbS/CdS Quantum dots | Visible red to near infrared | - | Exciplex state formation between perovskite and QDs | 1 - 3 | - | 0.01 | 31 |
| **ITO/PEDOT:PSS/PVK:PQD/Ca/Al or ITO/PEDOT:PSS/PVK/PFO:PQD/Ca/Al.** | M (3 layers) | Conjugated polymers: PVK and PFO | Blue region | (0.18, 0.1) to (0.2, 0.17) | FRET | 6 - 10 | 31 @ 9.5v | - | 32 |
| | | | Blue - Green | (0.18, 0.23) to (0.08, 0.52) | FRET | 6 - 16 | 277 @ 15v | | |
| **ITO/ PEDOT:PSS/ $MAPbBr_3$:POEA/ TPBi/ Ba/ Al** | M (3 layers) | 2D Perovskite | 532 – 462 nm | (0.21, 0.73) to (0.16, 0.1) | Adding amino head group of small molecules (POEA) to perovskite structure | - | 2146 @ 6.2v | 2.82 | 33 |
| **ITO/ PEDOT:PSS/ $PA_2CsPb_2I_7$/ TPBi/ Al** | M (3 layers) | Quasi 2D Perovskite | 654 – 691 nm | (0.708, 0.289) to (0.711, 0.277) | Quantum confinement effect in $PA_2CsPb_2I_7$ nanocrystals of various sizes | - | 189 @ 5.5v | 1.84 | 34 |
| **ITO/ $[Ru(bpy)_3](PF_6)_2$ : $[Os(phen)_3](PF_6)_2$ /Au** | Single layer | Ionic transition metal complexes (iTMC) | 610 – 690 nm | - | Polarization of $[Os(phen)_3]^{2+}$ inside $[Ru(bpy)_3]^{2+}(PF_6^-)_2$ leading to a change in the energy gap | 2.5 - 7 | 220 @ 3v | 0.75 | 35 |
| **ITO/ $[Ir(ppy)_2(dtb\text{-}bpy)](PF_6)$/Au** | Single layer | Iridium complex | Forward: 560nm Reverse: 580nm | Forward bias: (0.425, 0.549) Reverse bias: (0.491, 0.496) | Shift of the recombination zone | (-3) - (3) | 440 @ 3v | 5 | 36 |
| **ITO/[Ir(dfppz)2(dedaf)](PF6) : [Ir(ppy)2(biq)](PF6) : BMIMPF6/Ag** | Single layer | Iridium complexes | White region | (0.45, 0.40) (0.37, 0.39) (0.35, 0.39) | Host to guest energy transfer | 2.9-3.3 | 43 @3.3 | 4.0 | 37 |
| **Ag/Mg:Ag/TPP:$Alq_3$/α-NPD /ITO/PTCDA/α-NPD /Mg:Ag/$Alq_3$/Alq'$_2$OPh/α-NPD/ITO** | M (Tandem) | Organic small molecules | 450-670 nm | - | Independent operation of Red, Green, or Blue elements | 21-30 | 305 @ 0.2 mA | 0.96% | 38 |
| **ITO/PFN/TFB:MEH-PPV/$Cs_2CO_3$ /CuPC/NPB/TBADN/$Alq_3$/LiF/Al** | M (Tandem) | Conjugated polymers | - | Cool white: (0.33, 0.25) Warm white: (0.38, 0.39) Pure white: (0.35, 0.35) | Changing the concentration of MEH-PPV polymer in the active layer | Cool: 4 - 7 Warm: 5 - 9 Pure: 5 - 9 | 3500 @ 14v | - | 39 |
| **ITO/$Cs_2CO_3$/ MEH-PPV/$MoO_3$ /MWCNT+PEDOT:PSS/NPB/$Alq_3$/LiF/Al** | M (Tandem) | Conjugated polymers | - | (0.59, 0.38) – (0.47, 0.48) | Additional green emitting molecule ($Alq_3$) on top of subunit | 2 - 26 | 240 @ 7.5v | - | 40 |
| **Our Work** | Single layer | Iridium Complex | 518-610 nm Host FWHM: 8-10nm Guest FWHM: 35-40nm | (0.45, 0.47) – (0.25, 0.52) | Energy transfer from guest to host due to saturation of guest states | 3 - 6 | 10600 @ 4.3v | 3.81 | |

**Discussion of voltage-controlled color tunability.** Voltage-controlled color tunability (CT) has been achieved by various approaches (Table S2),[30-40] including reports of perovskites,[30-34] ionic transition metal complexes,[35-37] and multilayer architectures.[38-40] In most cases CT is achieved in complex multilayer LED structures that may contain bilayers of different color emitters and tandem structures in which different color LED sub-cells (e.g. blue and green or red) can be tuned by separate voltages.[38-40] This approach allows white color temperature fine tuning. Advantageous CT has been demonstrated in simple single layer architectures of LECs, [35-37] even leveraging the host-guest strategy with ionic transition metal complexes iTMCs to achieve improved efficiencies as high as 8 Lm/W.[35,37] However, these LECs exhibited low luminance (<220 cd/m$^2$) and modest CT spectral width. However using conventional hosts such as broad-band polymers or polyelectrolytes it was not possible to significantly increase brightness in these CT single layer LECs. We demonstrate in this work that, due to unique properties of perovskite hosts, they are a promising new landscape for bright and wide band CT in simplest architectures. This is because perovskites possess a unique combination of high mobility, exceptional luminescent quantum yields and inherent highly mobile internal ions that enable single layer LEC operation, surpassing the capabilities of previous host materials in host-guest systems. Having an ionic guest is particularly important for bright CT with a broad spectral band.

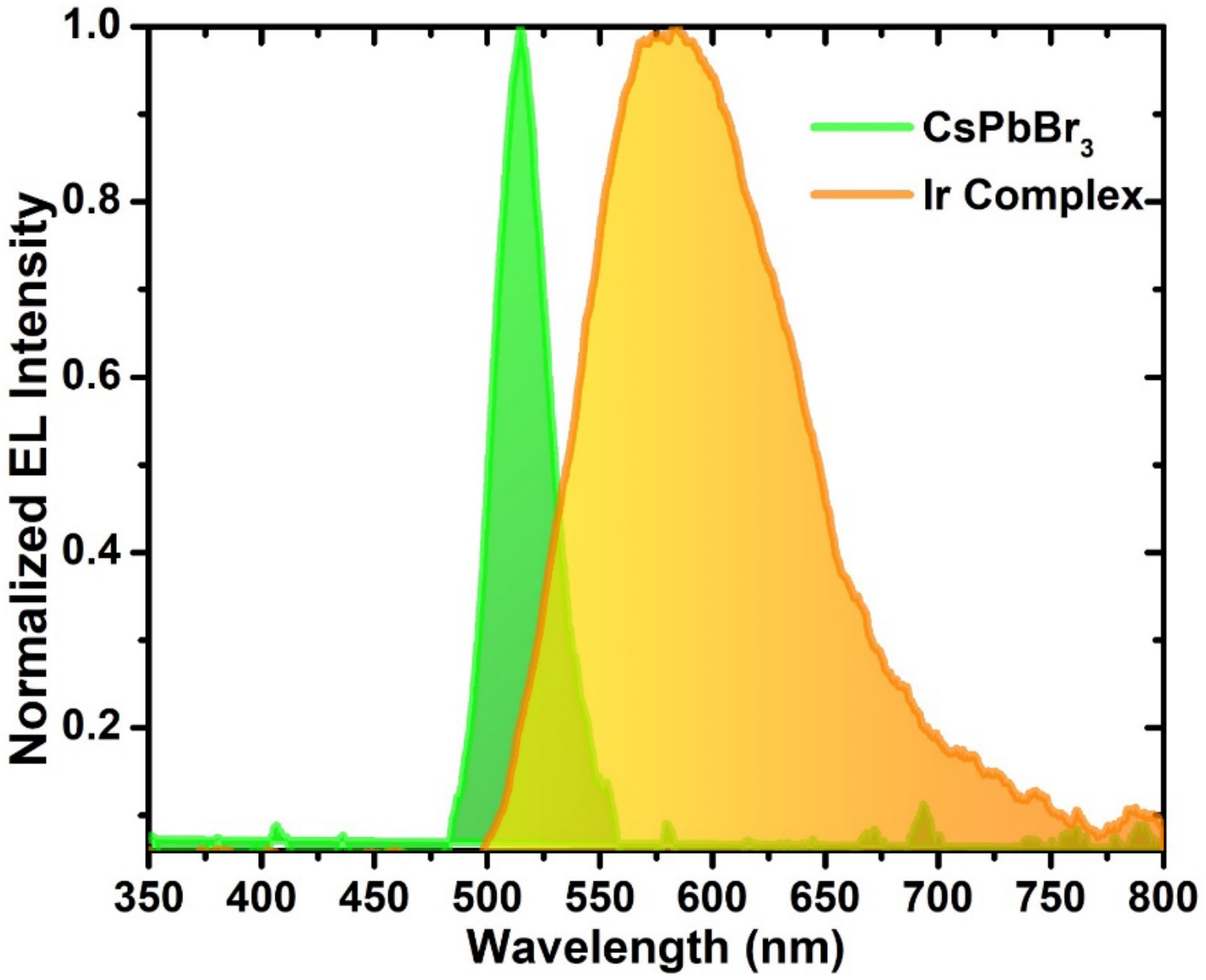

Figure S1: Electroluminescence (EL) spectra at 4.0 V for $CsPbBr_3$ + PEO (host and electrolyte) and [Ir(*t*-Bupmtr)$_2$(bathophen)][$PF_6$] (pure guest) PeLECs.

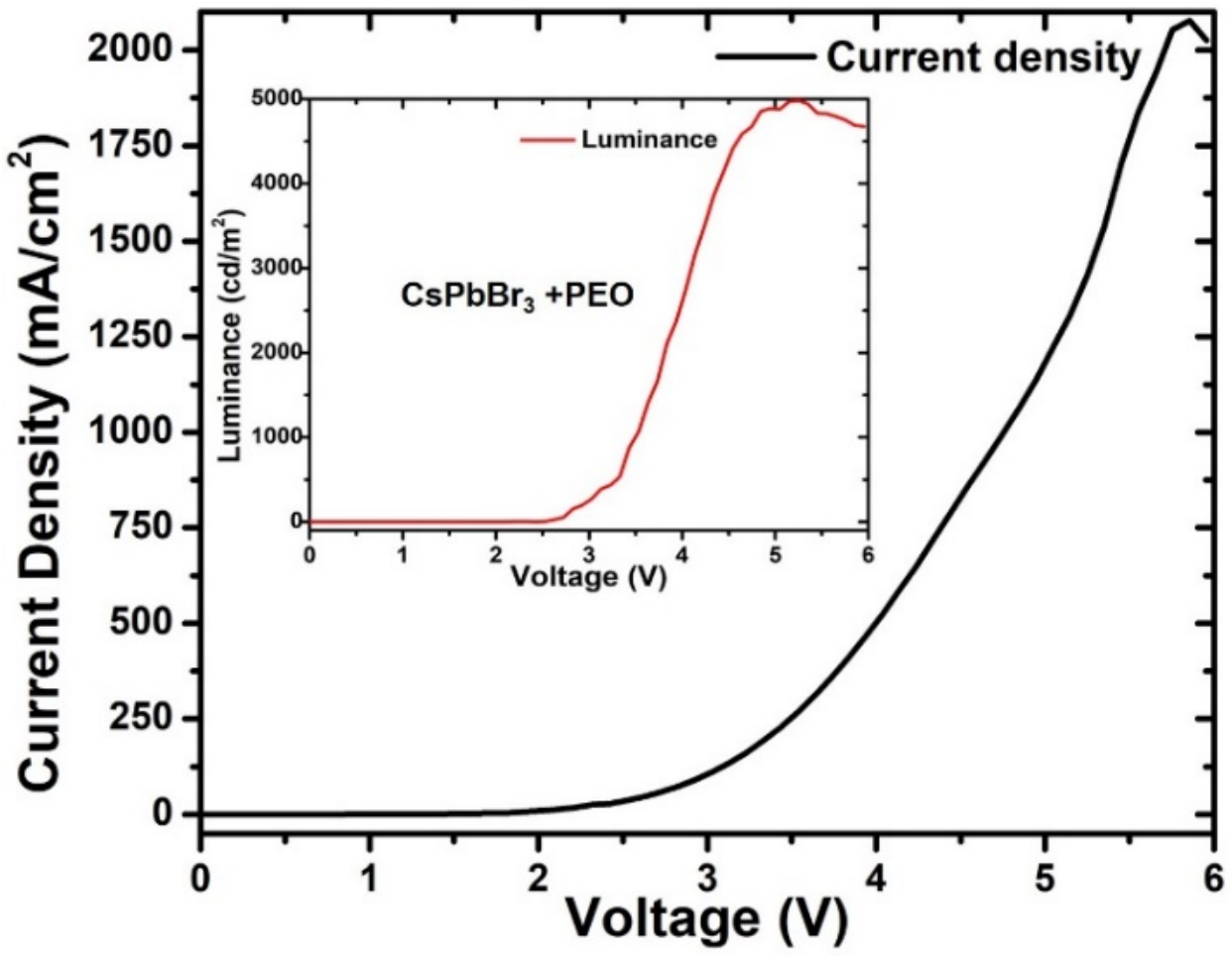

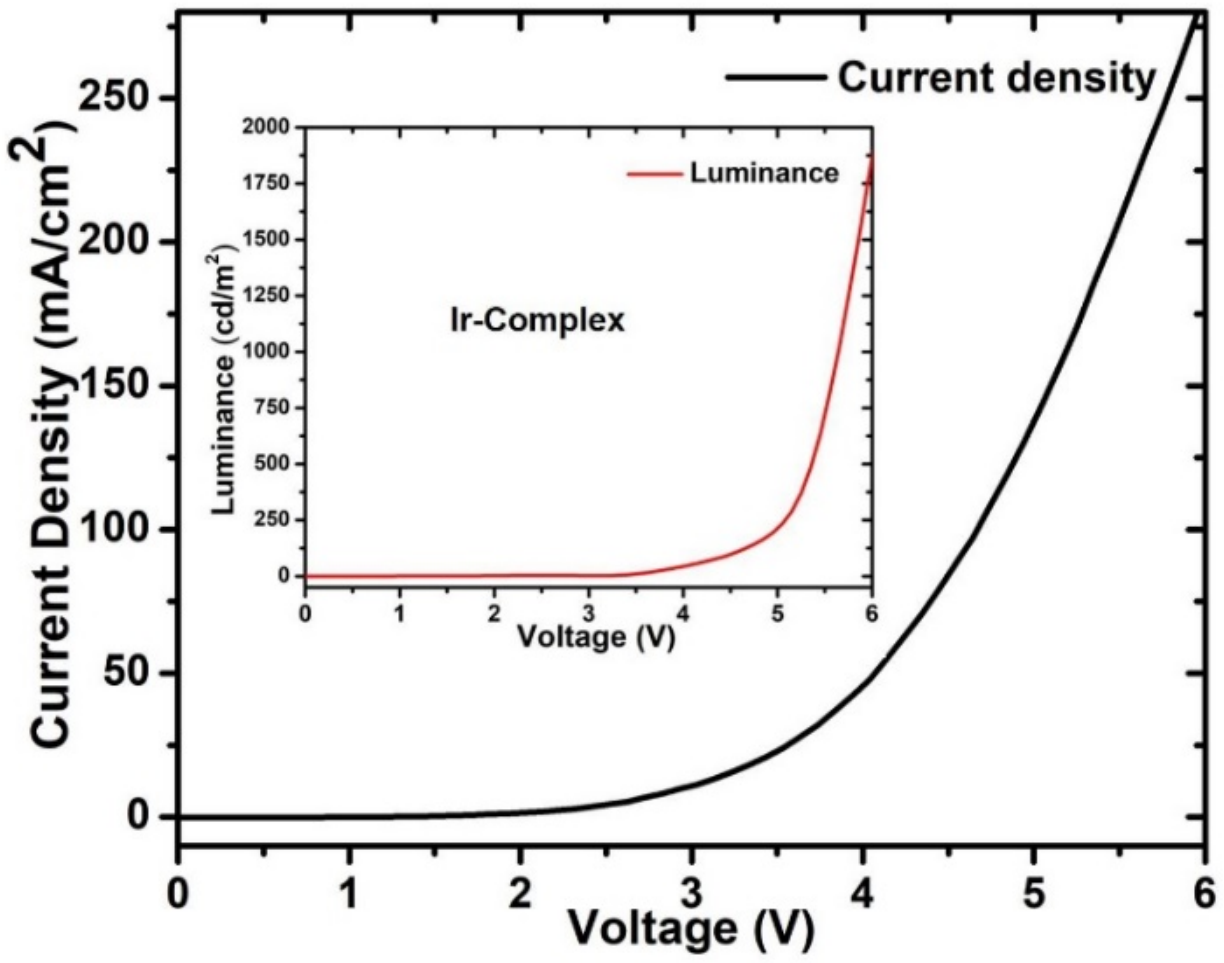

Figure S2: Current density versus voltage and luminance versus voltage (insets) for $CsPbBr_3$ + PEO and [Ir(*t*-Bupmtr)$_2$(bathophen)][$PF_6$] (pure guest) PeLECs.

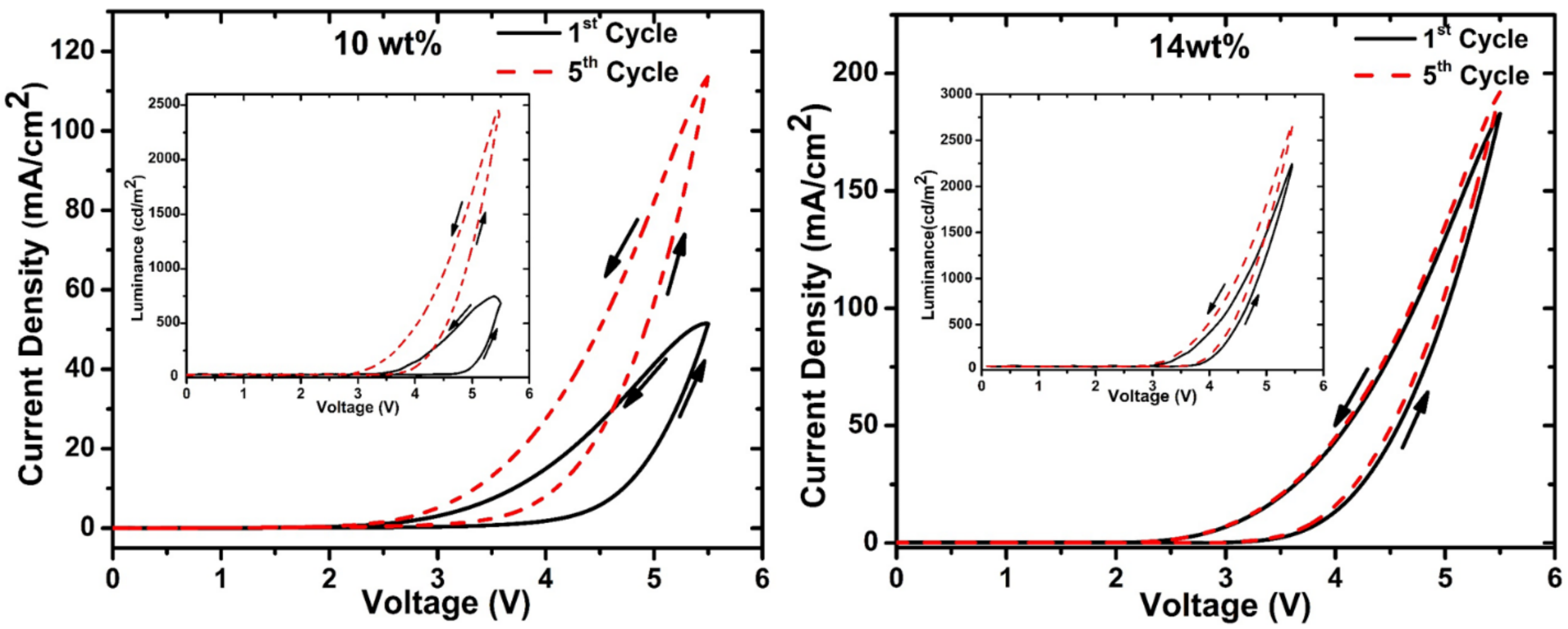

Figure S3: Current density versus voltage for host/guest PeLEC devices with 10 wt% or 14 wt% [Ir(*t*-Bupmtr)$_2$(bathophen)][$PF_6$] guest.

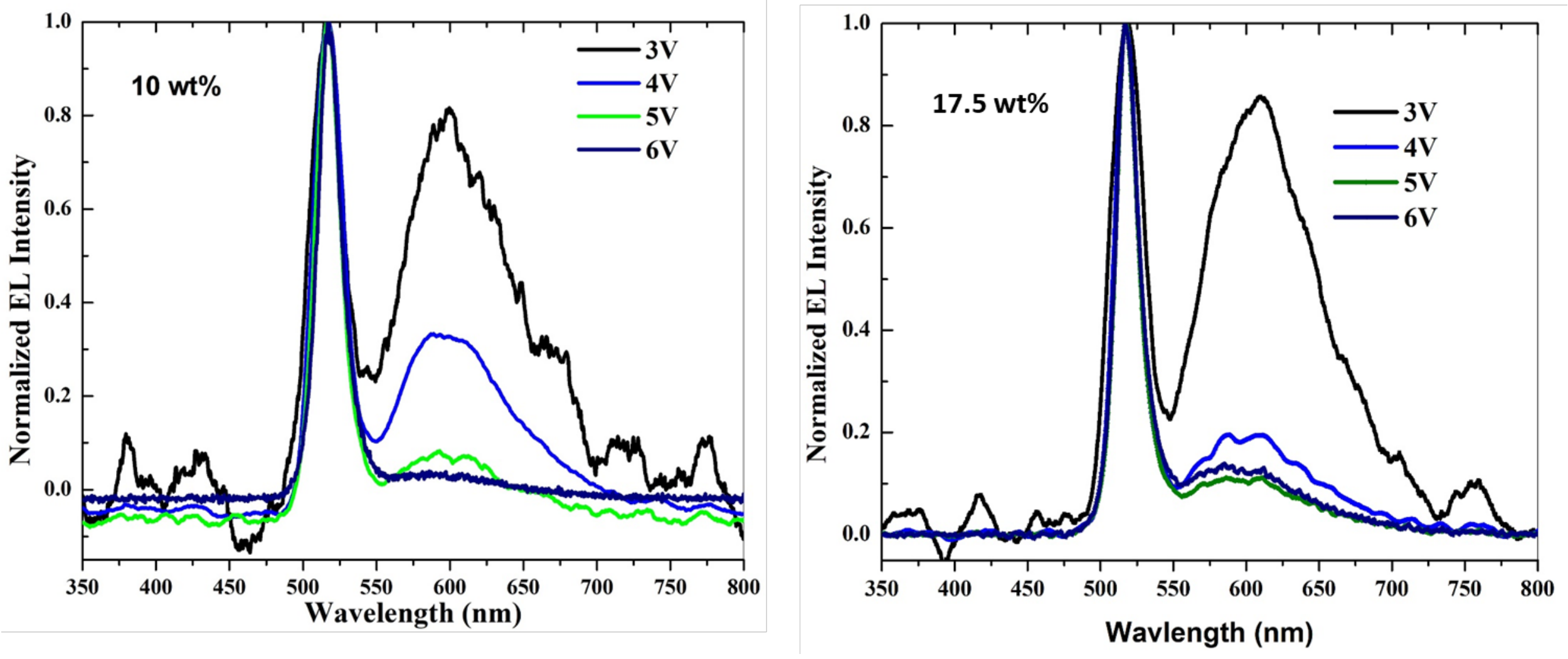

Figure S4: Electroluminescence spectra at various voltages for host/guest PeLEC devices with 10 wt% or 17.5 wt% [Ir(*t*-Bupmtr)$_2$(bathophen)][$PF_6$] guest.

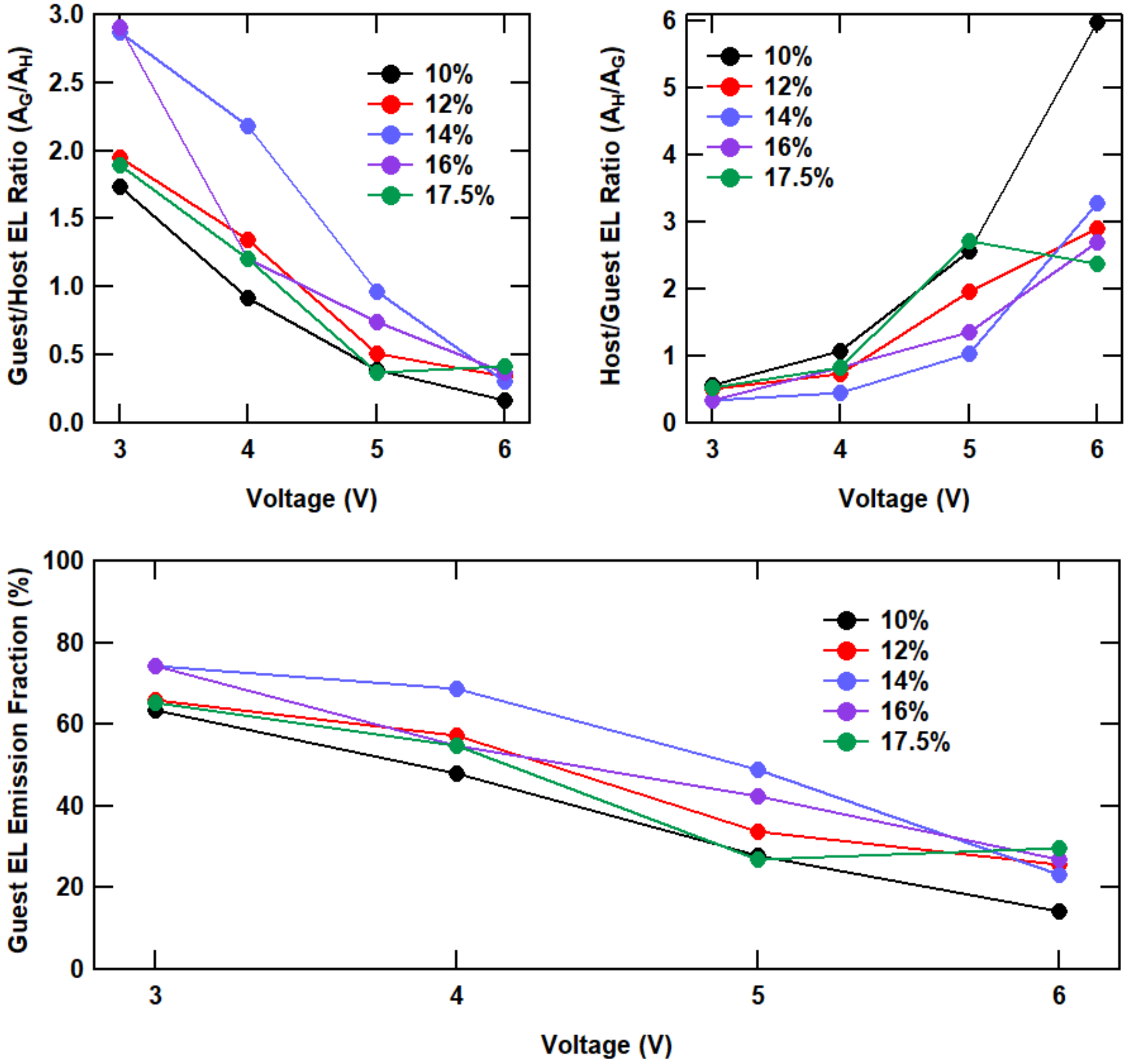

Figure S5: Guest/host and host/guest EL emission area ratios and guest EL emission fraction from the electroluminescence spectra for host-guest PeLECs with various weight fractions of guest at various applied voltages. Peak area ratios were calculated by integrating EL as a function of emission energy for each component peak.

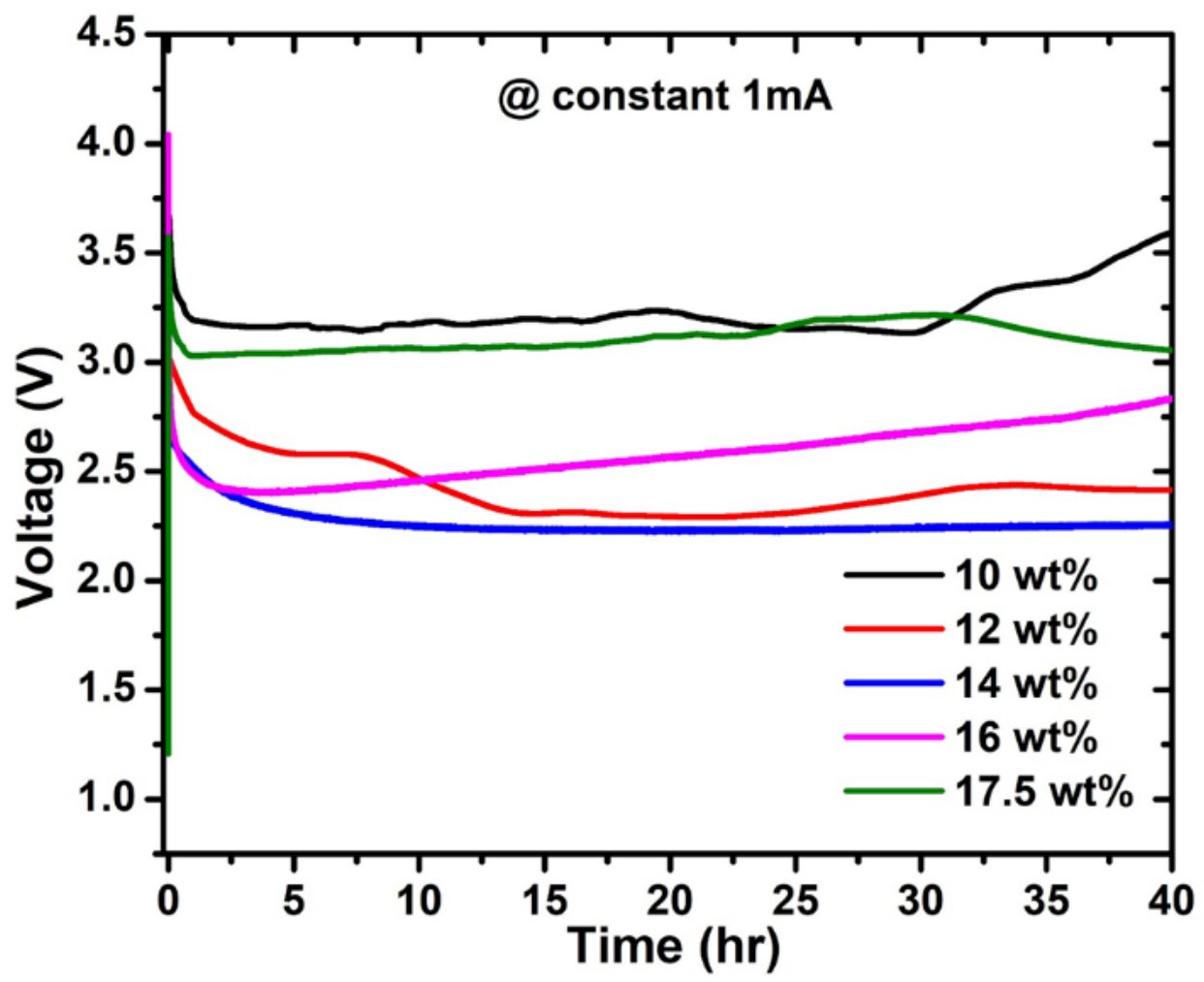

Figure S6: Voltage versus time for host/guest PeLEC devices with various weight fractions of guest operating at 1.0 mA constant current.

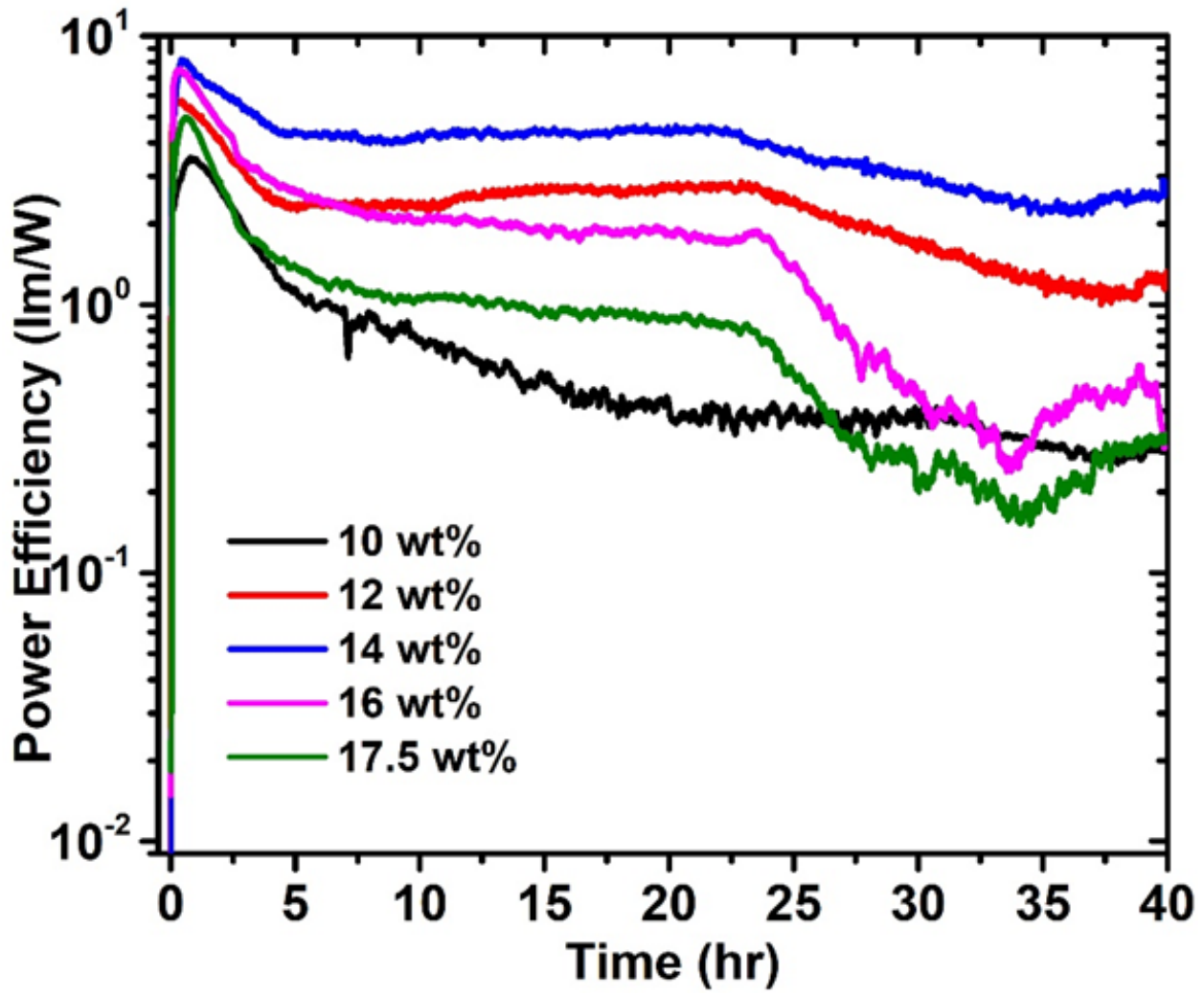

Figure S7: Power efficiency versus time for host/guest PeLEC devices with various weight fractions of guest operating at 1.0 mA constant current.

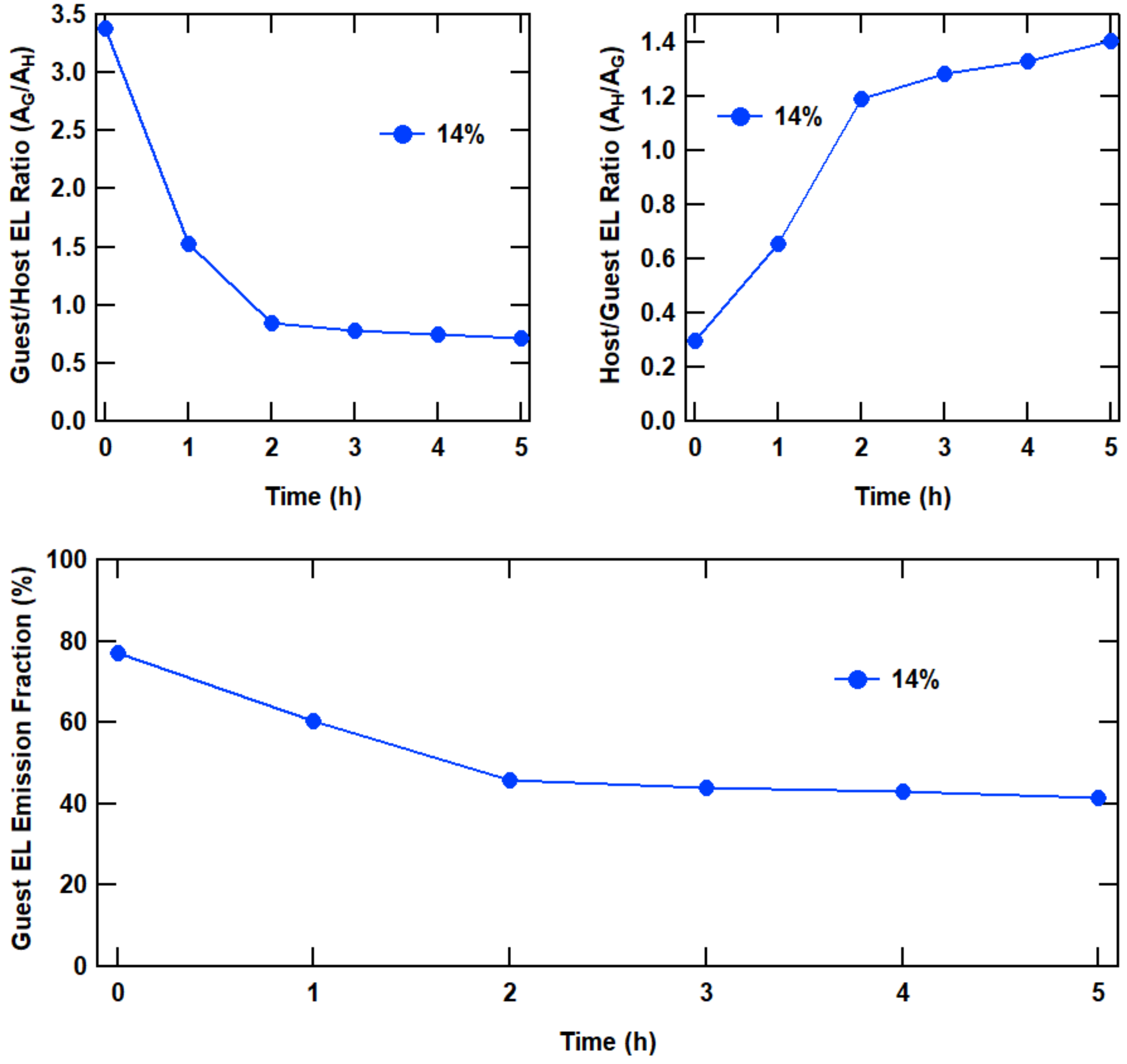

Figure S8: Guest/host and host/guest EL emission area ratios and guest EL emission fraction as a function of time from the electroluminescence spectra for host-guest PeLECs with 14 wt% guest. Peak area ratios were calculated by integrating EL as a function of emission energy for each component peak.

**Guest emitter design, synthesis, and characterization.** Heteroleptic $[Ir(C^N)_2(N^N)]^+$ complexes, where C^N is a cyclometallating ligand (e.g. 2-phenylpyridine, ppy) and N^N is a 1,2-diimine ligand such as 2,2'-bipyridine (bpy), are widely adopted due to synthetic accessibility of the ionic complex, long-lived triplet excited state, and tunable photophysical/photochemical properties. The frontier orbital structure of this type of complex consists of a HOMO residing over the phenyl rings of the cyclometallating ligands and the Ir *d* orbitals, while the LUMO is located on the ancillary ligand.[1] This spatial separation between the HOMO and LUMO allows independent orbital energy level tuning through functional group modifications; electron donating groups raise orbital energy levels while electron withdrawing groups lower orbital energy levels. In an effort to bathochromatically shift the absorption spectra, *t*-Bupmtr was utilized as an electron rich cyclometallating ligand to destabilize the HOMO, while bathophen, due to the increased π conjugation, was employed as the ancillary ligand to stabilize the LUMO. This chemical tuning of the bandgap was necessary to perform efficient energy harvesting from the perovskite, as host-guest blends with $[Ir(ppy)_2(bpy)][PF_6]$ derivatives[2] showed limited energy transfer.

Details of the synthesis procedure and structural characterization are presented in the Methods section, Supplementary Information Scheme 1 and Figures S10-S15. Figure 7a depicts the absorption and emission spectra of [Ir(*t*-Bupmtr)$_2$(bathophen)][PF$_6$]. Intense absorption peaks between 200 – 300 nm ($\varepsilon > 2*10^4$ $M^{-1}cm^{-1}$, Table 1) are assigned to spin-allowed intramolecular $^1\pi \rightarrow \pi^*$ transitions localized on the cyclometallating and ancillary ligands.[3] Transitions occurring above 300 nm, albeit with lower molar absorptivity ($\varepsilon < 2*10^4$ $M^{-1}cm^{-1}$, Table 1), are ascribed to intermolecular ligand-to-ligand charge transfer (LLCT) and metal-ligand-to-ligand charge transfer (MLLCT), with the latter typically controlled *via* HOMO/LUMO tuning.[1] Strong spin-orbit effects arising from the Ir(III) core allows spin-forbidden $^3$LLCT/$^3$MLLCT transitions to occur at a lower molar absorptivity than the corresponding [1]LLCT/[1]MLLCT.[1] The inset shows the absorption tail extending above 500 nm, directly resulting from the decreased HOMO/LUMO gap arising from ligand modifications. The structureless emission spectra, along with the long-lived deaerated excited state (Table S3), suggest emission arising from the charge separated, $^3$MLLCT excited state. The emission peak centered at 594 nm is comparatively red-shifted to other heteroleptic Ir(III) complexes previously formed.[4,5] Electrochemical stability, a key requisite for long-lived and robust OLED materials, was observed in the cyclic voltammogram (Fig. 1b), showing a reversible oxidation peak at 1.07 V and a reversible reduction at −1.52 V (Table 1). Ground state density functional theory (DFT) calculations (Fig. 1c) support assertions of a charge-separated HOMO/LUMO.

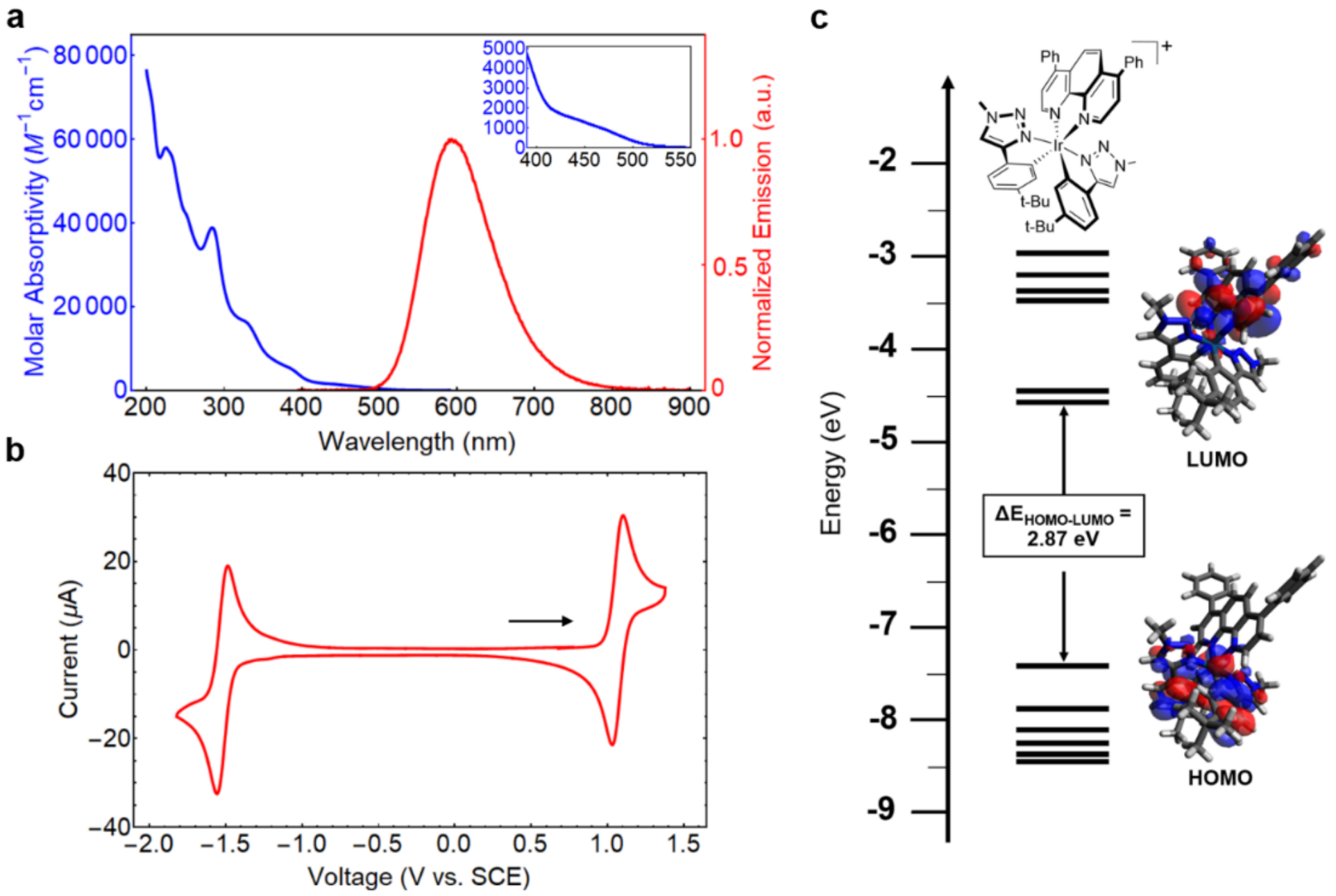

Figure S9. **Optical, electrochemical, structural, and electronic properties of the guest. a** Overlaid absorbance and emission spectra of $[Ir(t\text{-Bupmtr})_2(bathophen)]^+$. Absorbance spectra was measured with a 20 μM Acetonitrile solution at RT. Emission spectra was measured with a deaerated 0.5 mM Acetonitrile solution at RT. The inset shows the absorption tail above 400 nm. **b** Cyclic voltammogram (CV) of 1.0 mM Ir(III) solution with 0.1M tetrabutylammonium hexafluorophosphate supporting electrolyte in Acetonitrile; scan rate of 0.1 V/s. (**C**) Orbital energy level diagram of Ir(III) complex, showing the orbitals ranging from the HOMO – 5 level to the LUMO + 5 level. Spatially separated HOMO/LUMO orbitals are shown.

**Table S3 Photophysical and electrochemical properties of [Ir(*t*-Bupmtr)$_2$(bathophen)][$PF_6$].**

| $Abs_{max}$ (nm) ε ($10^4$ $M^{-1}cm^{-1}$) | $Em_{max}$ (nm) | $τ_o$[a] (μs) | $τ_{air}$[b] (μs) | $E_{ox}$ (V) | $E_{red}$ (V) |
|---|---|---|---|---|---|
| 224 (5.81), 253 (4.20), 284 (3.82), 328 (1.64), 384 (0.53), 466 (0.09) | 594 | 0.85 | 0.11 | 1.07 (73)[c] | -1.52 (64)[c] |

[a] Measured in a deaerated 0.5 mM Acetonitrile solution at RT. [b] Measured in an aerated 0.5 mM Acetonitrile solution at RT. [C] Redox peak separation (mV)

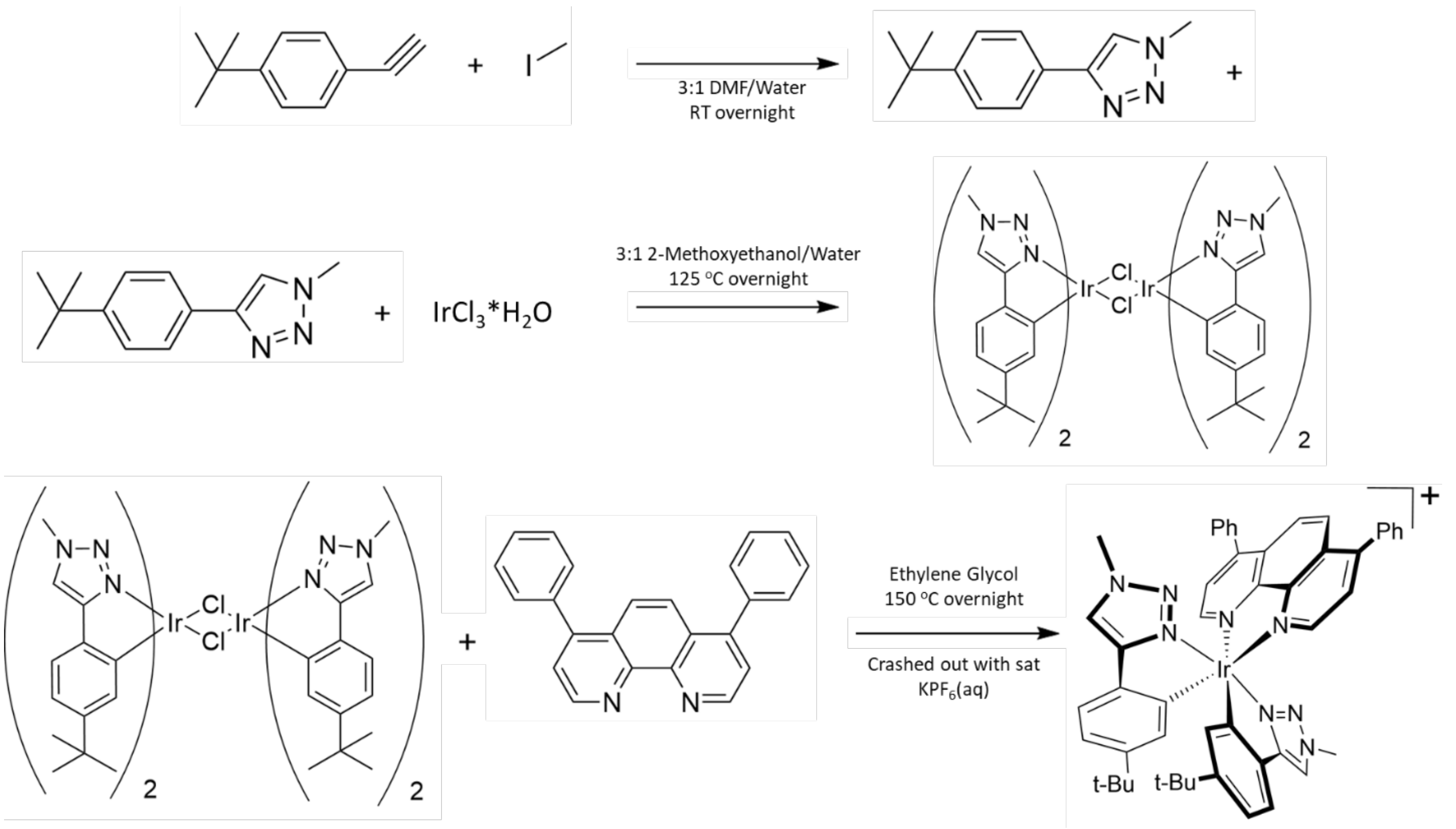

Scheme S1: Synthesis of [Ir($t$-Bupmtr)$_2$(bathophen)][$PF_6$]

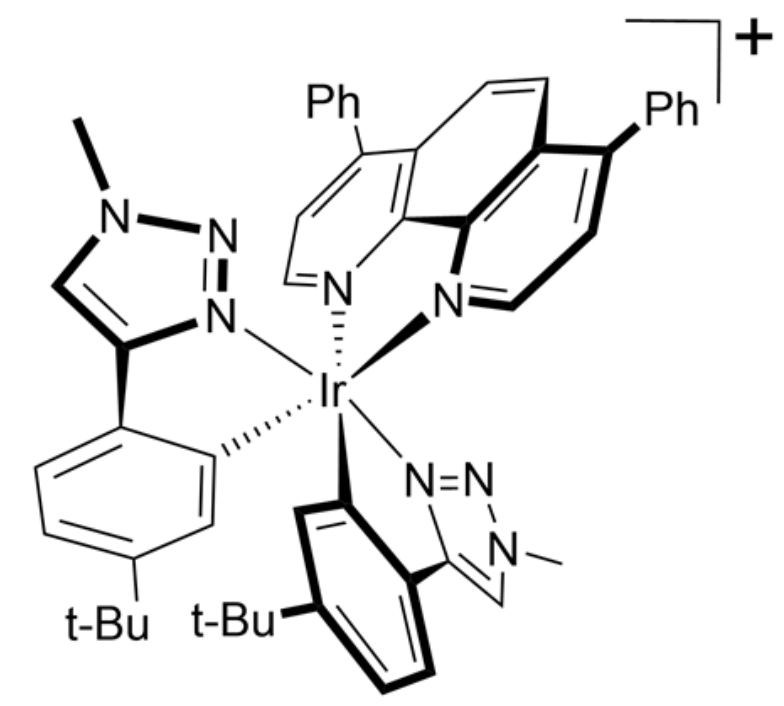

$^{1}H$ NMR

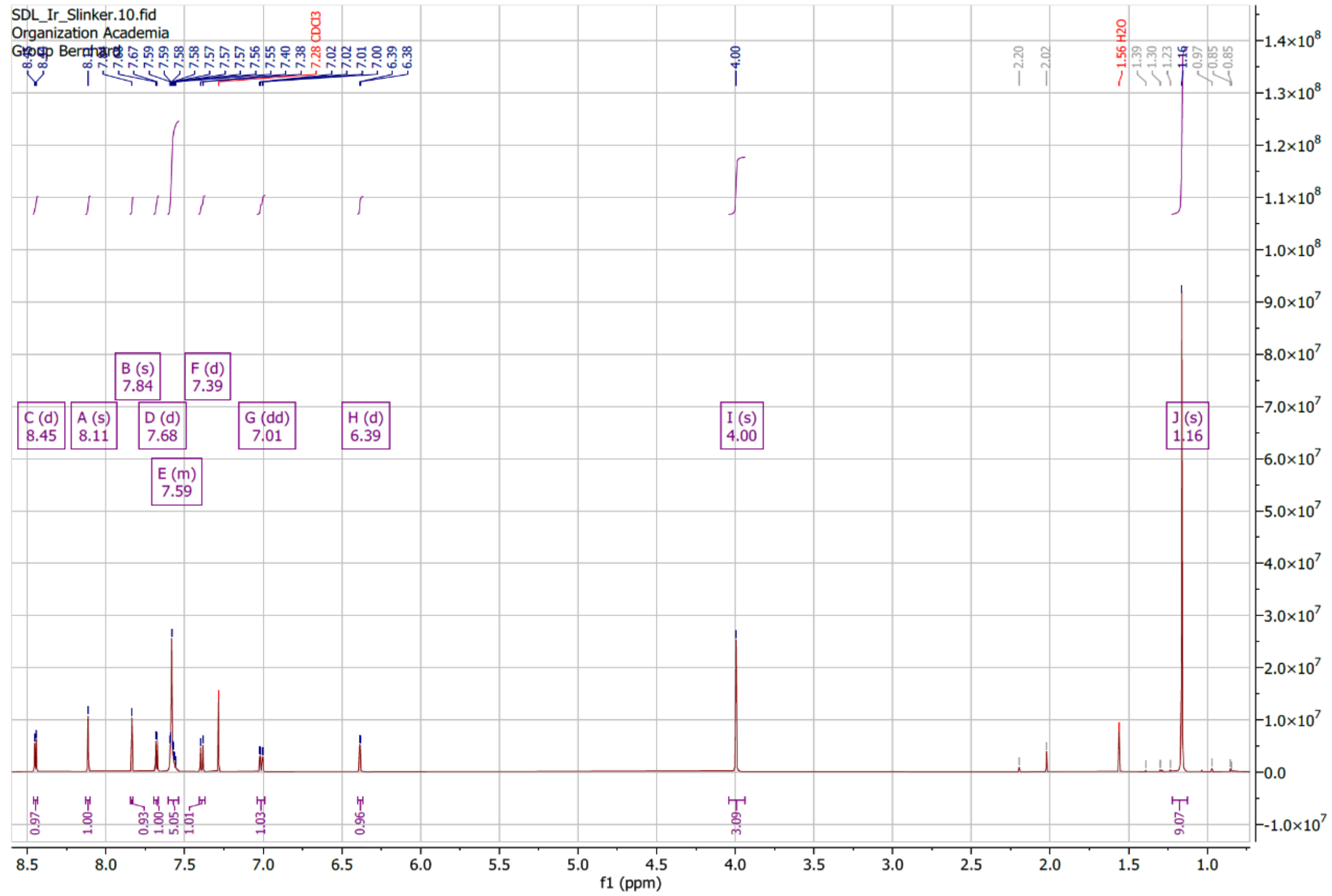

Figure S10: $^{1}H$ NMR for $[Ir(t\text{-}Bupmtr)_2(bathophen)][PF_6]$

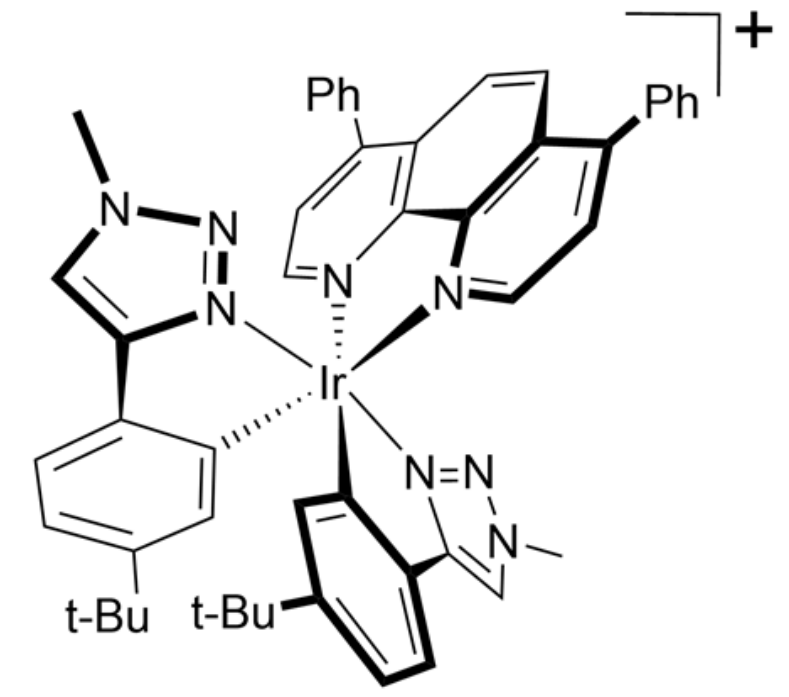

$^{13}C$ NMR

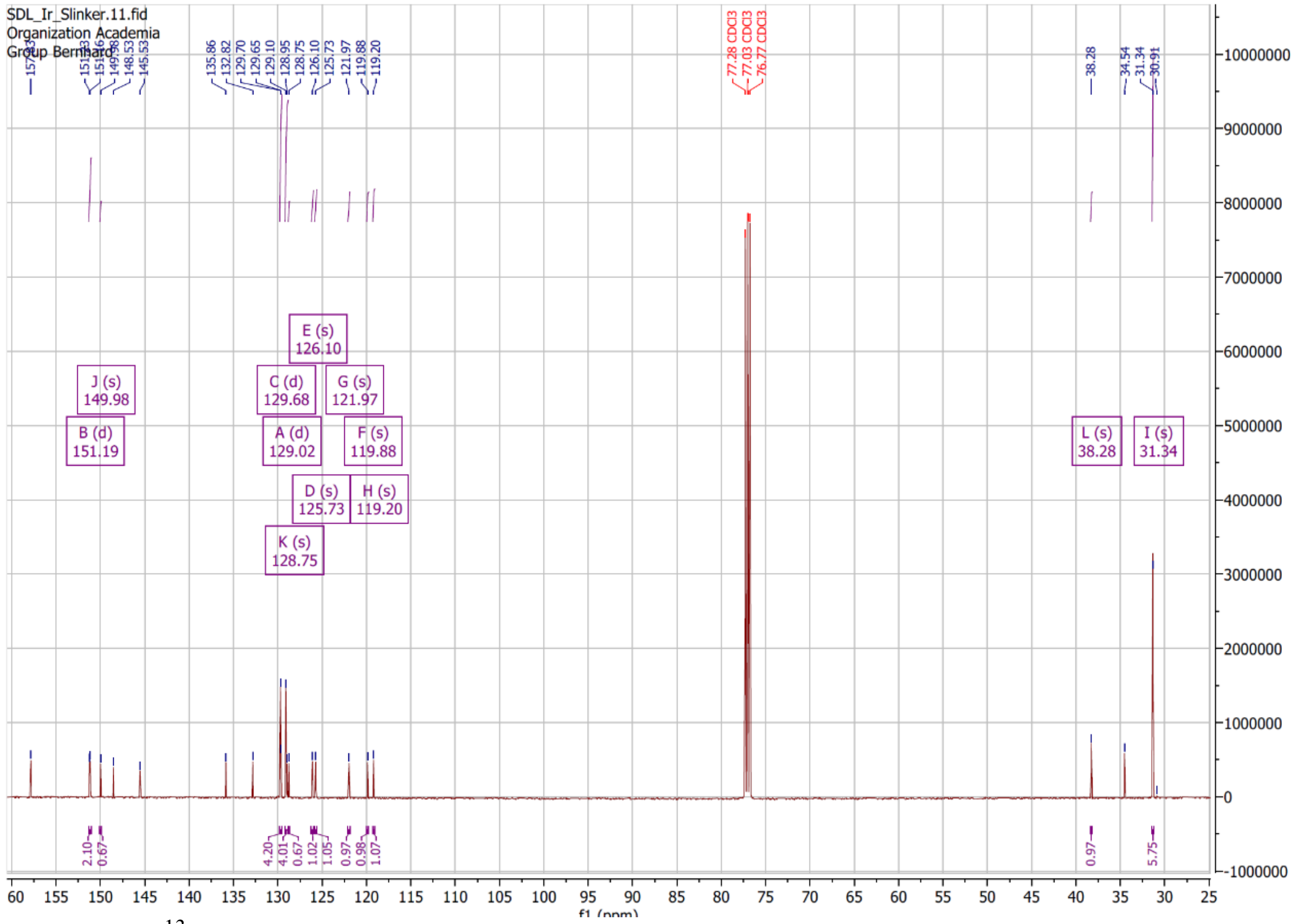

Figure S11: $^{13}C$ NMR for [Ir(*t*-Bupmtr)$_2$(bathophen)][$PF_6$].

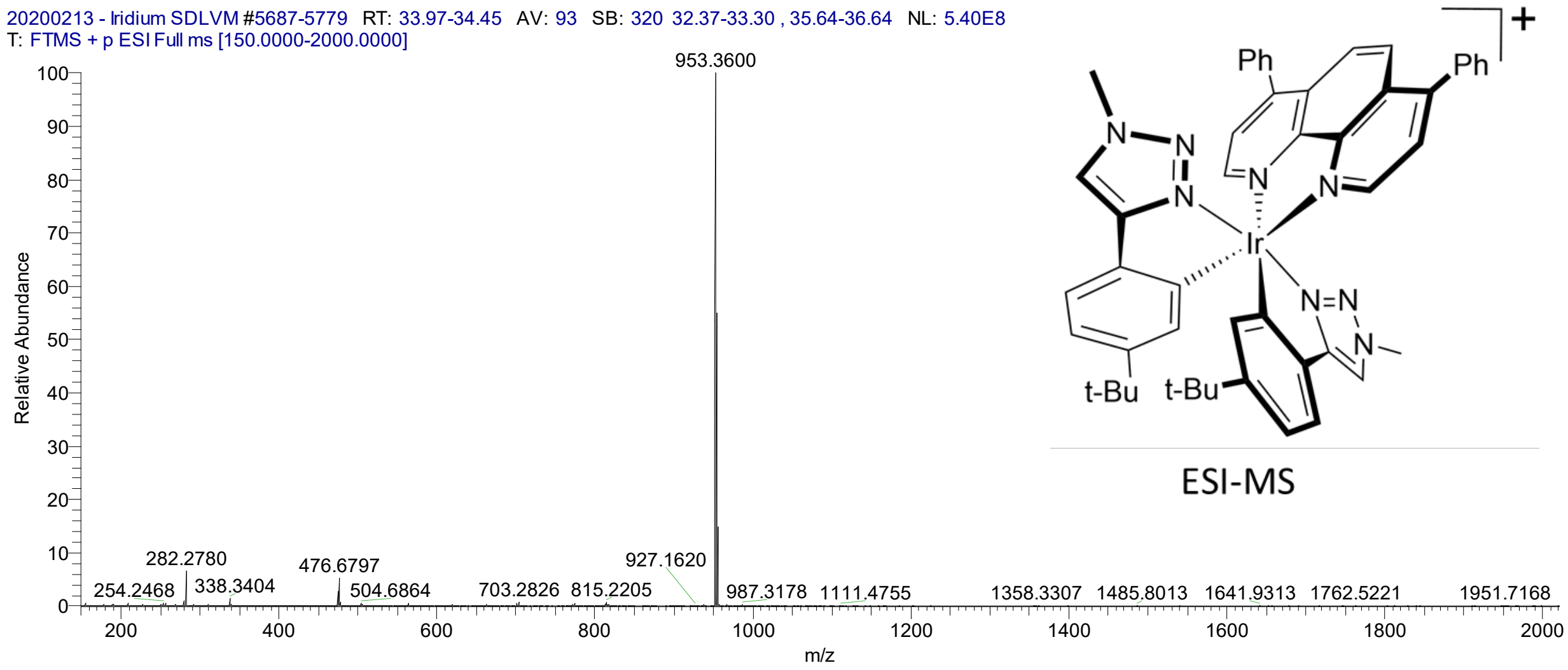

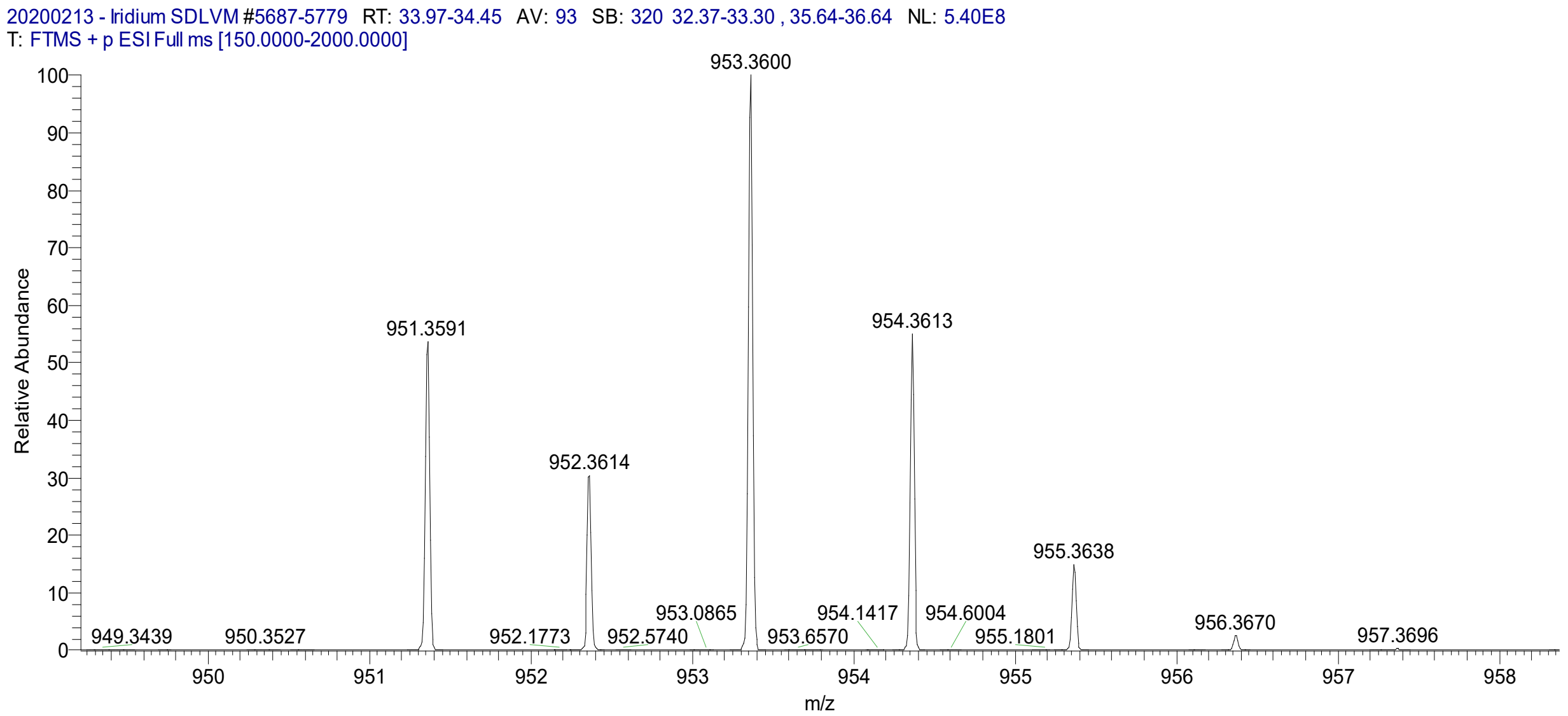

Figure S12: ESI-MS for [Ir($t$-Bupmtr)$_2$(bathophen)][$PF_6$].

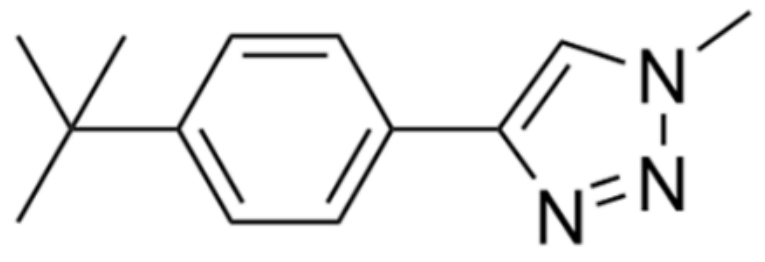

$^1$H NMR

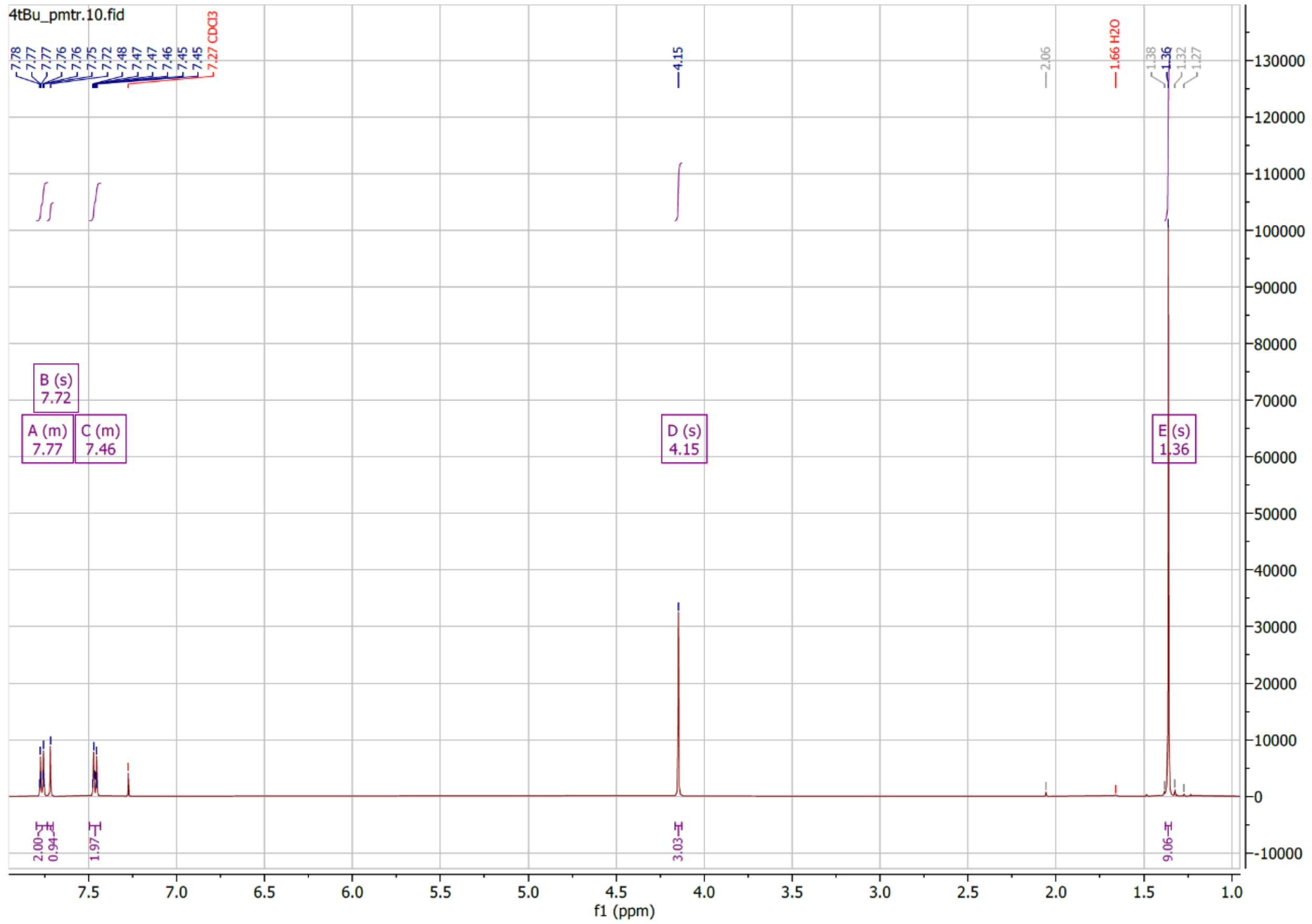

Figure S13: $^1$H NMR for the *t*-Bupmtr ligand.

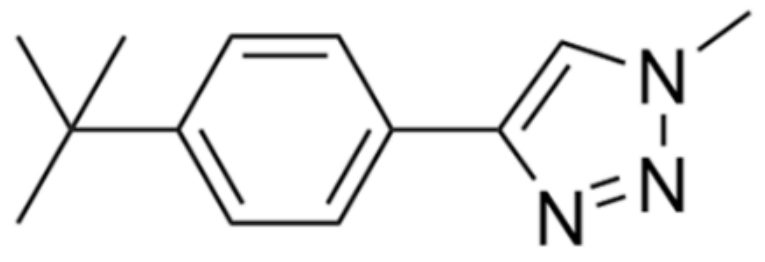

$^{13}$C NMR

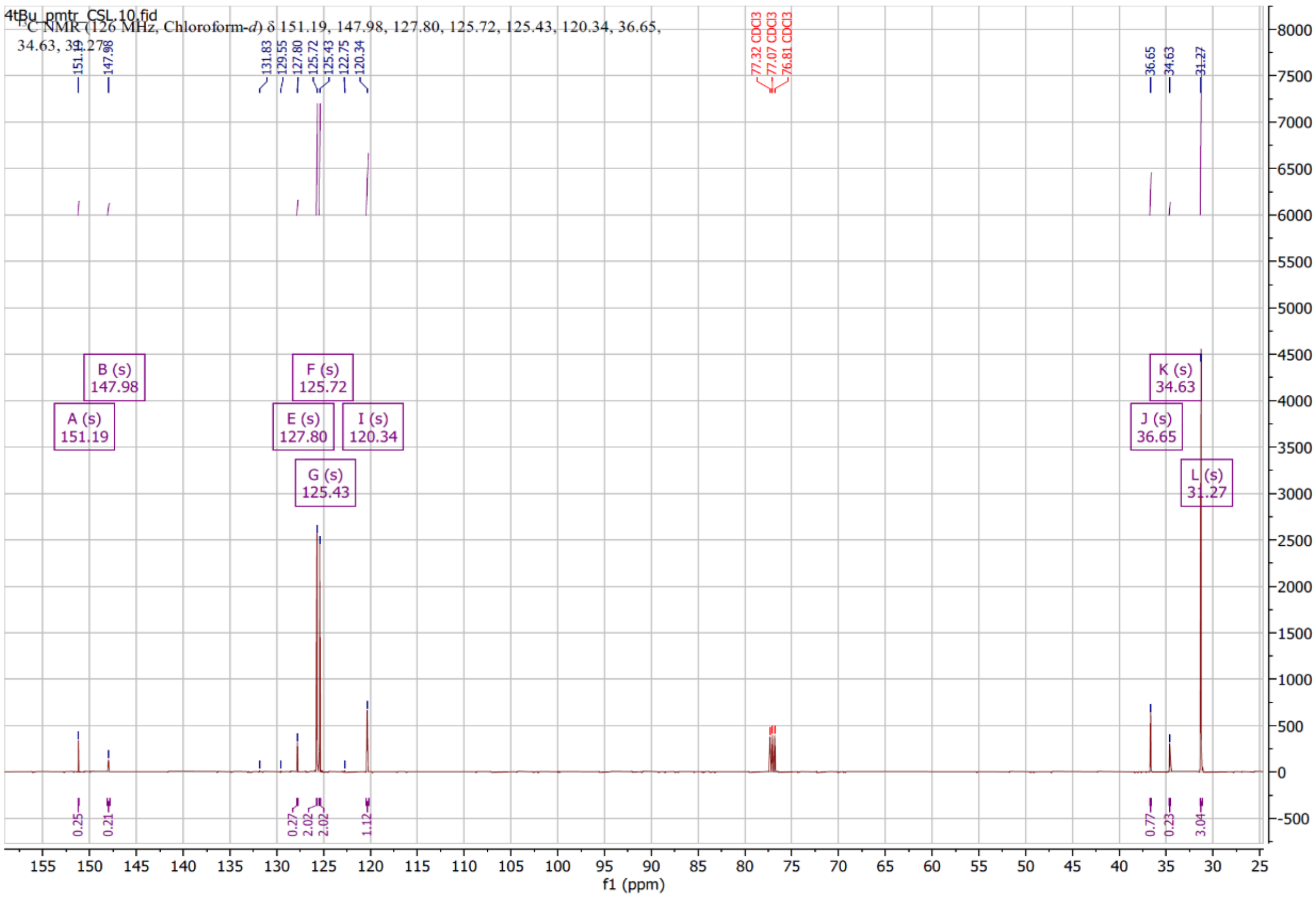

Figure S14: $^{13}$C NMR for the *t*-Bupmtr ligand.

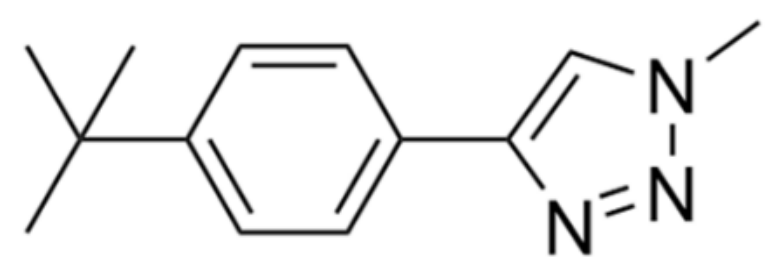

ESI-MS

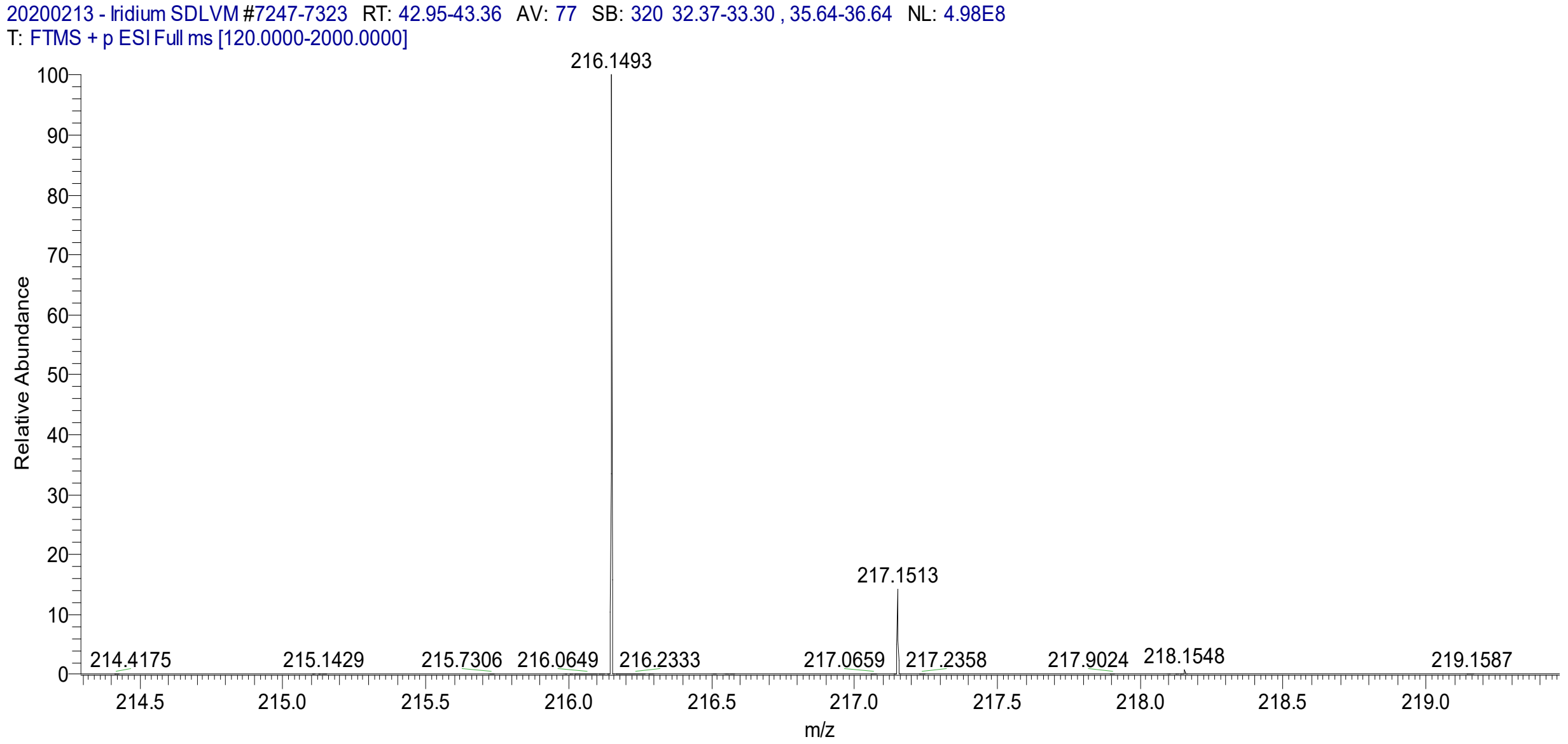

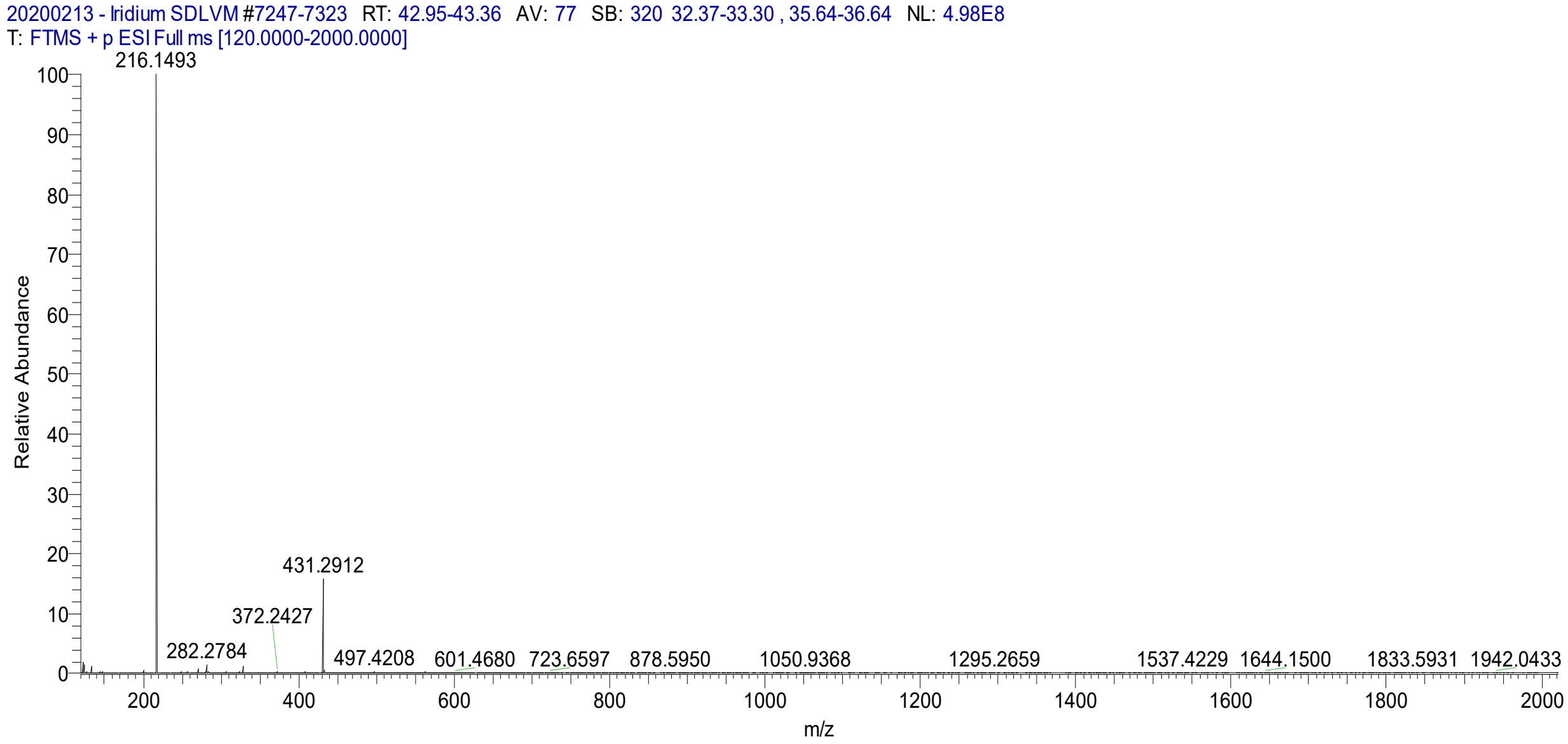

Figure S15: ESI-MS for the *t*-Bupmtr ligand.